\newif{\ifjournal}
  \journaltrue

\ifjournal
  \documentclass{mn2e}
  \usepackage{graphicx,mathptm}
  \newcommand{\gtrsim}{\ga} 
  \newcommand{\lesssim}{\la} 
\else
  \documentclass{paper}
  \newcommand{\ga}{\gtrsim} 
  \newcommand{\la}{\lesssim} 
\fi

\renewcommand{\d}{\mathrm{d}}

\begin{document}  
  
\title{cD galaxy contribution to the strong lensing cross sections of
  galaxy clusters}
 
\ifjournal\author[Meneghetti et al.]
  {Massimo Meneghetti$^{1,2}$, Matthias Bartelmann$^2$, Lauro
   Moscardini$^1$\\
   $^1$Dipartimento di Astronomia, Universit\`a di Padova, 
   vicolo dell'Osservatorio 2, I--35122 Padova, Italy\\ 
   $^2$Max-Planck-Institut f\"ur Astrophysik,
   Karl-Schwarzschild-Strasse 1, D--85748 Garching, Germany} 
\else\author
  {Massimo Meneghetti$^{1,2}$, Matthias Bartelmann$^2$, Lauro
   Moscardini$^1$\\
   $^1$Dipartimento di Astronomia, Universit\`a di Padova, 
   vicolo dell'Osservatorio 2, I--35122 Padova, Italy\\ 
   $^2$Max-Planck-Institut f\"ur Astrophysik,
   Karl-Schwarzschild-Strasse 1, D--85748 Garching, Germany} 
\fi

\date{Accepted 2001 ???? ???; Received 2000 ???? ???;  
in original form 2000 ???? ??}  
  
\ifjournal\maketitle\fi
  
\begin{abstract}
We perform ray-tracing simulations evaluating the effect of a cD
galaxy on the strong lensing properties of five galaxy cluster halos
obtained from N-body simulations. The cD galaxy is modelled using both
axially symmetric and elliptical models and assuming several masses
for its dark matter halo. The effect of the cD orientation with
respect to the mass distribution of the host galaxy cluster is also
investigated. The simulations are carried out in an open and a flat
model universe with cosmological constant. We find that the
enhancement of the cluster lensing cross sections for long and thin
arcs due to the presence of a massive cD at the cluster centre is
typically less than $100\%$, depending on the model used for the cD
galaxy and its orientation. The impact of the cD on the cluster
efficiency for producing radially magnified images is larger only for
those clusters whose lensing cross section for radial arcs is very
small in absence of the central galaxy.  We conclude that the presence
of a cD galaxy at the cluster centre can only moderately influence the
cluster efficiency for strong lensing and in particular fails to
explain the discrepancy between the observed number of giant arcs in
galaxy clusters and their abundance predicted from lensing simulations
in the currently most favoured $\Lambda$CDM model.
\end{abstract}

\ifjournal\else\maketitle\fi

\section{Introduction}

Many authors have pointed out that the observed abundance of
gravitational arcs whose length-to-width ratio exceeds a given
threshold can be used for constraining the cosmological model and its
parameters. Indeed, the occurrence of those rare events caused by
highly non-linear effects in cluster cores is determined by several
factors related to the geometry of the universe, as well as to the
abundance, evolution and internal structure of the lensing clusters,
which all depend on cosmology.

Investigating the lensing properties of a large sample of numerically
modelled galaxy clusters using the ray-tracing technique, Bartelmann
et al.~(1998) showed that the expected number of \emph{giant} arcs,
usually defined as arcs with length-to-width ratio exceeding 10 and
apparent $B$-magnitude less than $22.5$ \cite{wu93}, changes by orders
of magnitude between different cosmological models. In particular,
they found that the expected number of giant arcs is smaller by a
factor of ten in a flat, low density universe with a matter density
parameter $\Omega_0=0.3$ and a cosmological constant
$\Omega_\Lambda=0.7$ ($\Lambda$CDM) than in an open, low-density
cosmological model with $\Omega_0=0.3$ and $\Omega_\Lambda=0$ (OCDM).

Observations of the abundance of gravitational arcs in galaxy clusters
seem to be consistent only with the predictions for an open universe
\cite{bartelmann98,lefevre94,gioia94,luppino99,gladders03,zaritsky03}.
This is in pronounced disagreement with other observational results,
in particular those obtained from the recent experiments on the cosmic
microwave background \cite{debernardis02} and the observations of
high-redshift type-Ia supernovae
\cite{perlmutter97,perlmutter98}, which all suggest
instead that the cosmological model most favoured by the data is
spatially flat and dominated by a cosmological constant.

Several studies tried to resolve this inconsistency, but satisfactory
explanations remain to be found. Similar estimates of the number of
long and thin arcs were attempted adopting isothermal analytic instead
of numerical cluster lens models \cite{cooray99,kaufmann00}. The
results of these studies differ from those of Bartelmann et al.~(1998)
in that they are almost insensitive to the cosmological constant and
predict similar abundances of arcs in the $\Lambda$CDM and in the OCDM
models. However, Meneghetti, Bartelmann \& Moscardini (2002) recently
showed that analytic models are inadequate for quantitative studies of
arc statistics, because asymmetries and substructures in the deflector
play a relevant role for determining the galaxy cluster efficiency for
strong lensing and cannot properly be taken into account when using simple
analytic models for modelling cluster lenses. Moreover, Bartelmann et
al.~(2003) demonstrated that elliptical cluster models with the
density profile found in numerical simulations by Navarro et
al.~(1996) were able to reproduce the relative change of an order of
magnitude in the arc cross sections produced by clusters in the
$\Lambda$CDM and OCDM models.

By investigating the properties of giant gravitational arcs in the
Red-Sequence Cluster Survey, Gladders et al.~(2003) found a surprisingly
high frequency of arc systems in high redshift clusters ($z>
0.64$). Their determination of lensing incidence is in agreement with
the results by Zaritsky \& Gonzalez~(2003), who measured the abundance
of strong gravitational lensing events in the Las Campanas Distant Cluster
Survey. As the authors of these studies suggest, some property
associated with clusters at early times might result in a higher
efficiency for strong lensing. Torri et al.~(2003) have recently
studied the effects on strong cluster-lensing cross sections induced
by major merger events within the clusters. Their results indicate
that the cross section for giant arcs can change by almost one order
of magnitude while a massive substructure crosses the cluster
centre. Numerical simulations show that mergers are frequent events in
clusters. Therefore, they may offer an explanation as to why
so many arc systems are observed in both high and low redshift
clusters. Further investigations into the impact of mergers on arc
statistics are now in progress.

Meneghetti et al.~(2001) also investigated the effect of cluster
galaxies on arc statistics but found that they do not introduce
perturbations strong enough for significantly changing the number of
arcs and the distributions of lengths, widths, curvature radii and
length-to-width ratios of long arcs.

However, the possible effects due to the presence of a massive cD
galaxy near the cluster centre were not investigated there. The
centres of massive galaxy clusters are generally dominated by such
very massive ($\sim10^{13}\,M_\odot$) galaxies, which could in
principle noticeably affect the strong-lensing properties of their
host clusters. In fact, due to their more concentrated dark matter
halos, they may steepen the inner slope of the cluster density profile
and push the cluster critical curves to larger distances from the
centre. Thus, the length of the critical curves may be increased, and
thus the probability for long arcs to form. Moreover, cD galaxies may
help their host clusters to reach the critical central surface density
for producing critical curves and becoming efficient strong lenses.

Understanding the effect of massive central galaxies on the strong
lensing properties of galaxy clusters is an important issue also
because several studies attempted using arc statistics for
constraining the density profiles of lenses or the nature of the dark
matter composing their halos
\cite{wu96,molikawa99,oguri01,molikawa01,oguri02,meneghetti01}. The
conclusions from all these studies may change should the effect of a
central massive galaxy turn out to be relevant.

Thus, we investigate in this paper whether the presence of a massive
cD galaxy at the centre of a galaxy cluster can significantly change
the cluster's ability to produce strong lensing events and reconcile
the lensing cross section of $\Lambda$CDM clusters with the
observations. For doing so, we study the lensing properties of a
sample of five galaxy clusters numerically simulated in both the
$\Lambda$CDM and the OCDM cosmological models, and we measure their
efficiency for producing tangential and radial arcs before and after
the inclusion of a cD galaxy. The central galaxy is modelled using
both axially symmetric and elliptical models and assuming a range of
virial masses and possible orientations with respect to the mass
distribution of the host cluster.

The plan of the paper is as follows. In
Section~\ref{section:expectations}, we use simple analytic models for
qualitatively estimating the impact of a cD galaxy on the strong
lensing properties of galaxy clusters. In
Section~\ref{section:nummod}, we describe the numerical models used
and the methods adopted for the lensing simulations. The results are
discussed in Section~\ref{section:results}. Finally, conclusions are
summarised in Section~\ref{section:conclusions}.

\section{Expectations}
\label{section:expectations}

We begin by investigating with the help of analytical models what
impact we can expect from cD galaxies on the strong-lensing properties
of galaxy clusters. We will use two different analytic lens models,
one with a singular isothermal density profile, and the other with the
density profile found in numerical simulations by Navarro et
al.~(1996; hereafter NFW).

\subsection{Axially symmetric models}

The radial density profile of a singular isothermal sphere (hereafter
SIS) is given by
\begin{equation}
  \rho(r)=\frac{\sigma_v^2}{2 \pi G r^2}\;,
\end{equation}
where $\sigma_v$ is the velocity dispersion of the halo. The NFW
density profile is instead given by
\begin{equation}
  \rho(r)=\frac{\rho_\mathrm{s}}
  {(r/r_\mathrm{s})(1+r/r_\mathrm{s})^2}\;,
\end{equation}
where $\rho_\mathrm{s}$ and $r_\mathrm{s}$ are characteristic density
and distance scales, respectively (see Navarro et al.~1997). These two
parameters are not independent. The ratio between $r_\mathrm{s}$ and
the radius $r_{200}$ within which the mean halo density is $200$ times
the critical density is called concentration,
$c=r_{200}/r_\mathrm{s}$. As results from numerical simulations show,
the concentration parameter $c$ can be expressed as a function of the
halo virial mass, which thus is the only free parameter. The
concentration also depends on the cosmological model, implying that
the lensing properties of haloes with identical mass are different in
different cosmological models if they are modelled as NFW spheres.
The NFW halo profile falls off steeper than isothermal at radii beyond
$r_\mathrm{s}$, but flattens towards the halo centre. These different
features lead to markedly different lensing properties of the NFW
compared to the SIS model (see the discussions in Perrotta et al.~2002
and Meneghetti et al.~2003).

Several algorithms were developed for computing the concentration
parameter from the halo virial mass
\cite{navarro97,bullock01,eke01}. They are all based on the findings
that numerically simulated haloes tend to be more concentrated the
earlier they form, and that their central density reflects the mean
cosmic density at their formation time. Since haloes of lower mass
form earlier in hierarchical models than haloes of higher mass, the
concentration is a decreasing function of the halo mass.

In this work, we adopt the algorithm proposed by Navarro et
al.~(1997), which first assigns to a halo of mass $M$ a collapse
redshift $z_\mathrm{coll}$ defined as the redshift at which half of
the final mass is contained in progenitors more massive than a
fraction $f$ of the final mass. Then the density scale of the halo is
assumed to be some factor $C$ times the mean cosmic density at the
collapse redshift. They recommend setting $f=0.01$ and $C=3\times
10^3$ because their numerically determined halo concentrations were
well fit adopting these values.

In the case of axially symmetric models the computation of the
deflection angles reduces to a one-dimensional problem. We define the
optical axis as the straight line passing through the observer and the
lens centre and introduce the physical distances perpendicular to the
optical axis on the lens and source planes, $\xi$ and $\eta$,
respectively. We then fix a length scale $\xi_0$ on the lens plane and
define the dimensionless distance $x\equiv\xi/\xi_0$ from the lens
centre. By projecting $\xi_0$ on the source plane, we define a
corresponding length scale $\eta_0=\xi_0\,(D_\mathrm{s}/D_\mathrm{l})$
on the source plane, where $D_\mathrm{s}$ and $D_\mathrm{l}$ are the
angular diameter distances to the source and lens planes,
respectively. Like on the lens plane, we define a dimensionless
distance from the optical axis $y\equiv\eta/\eta_0$ on the source
plane.

Using this dimensionless formalism, the lensing potential of a SIS
lens can be written as
\begin{equation}
  \psi(x)=|x|\;,
\label{eq:sisPsi}
\end{equation} 
if
$\xi_0=4\pi(\sigma_v/c)^2\,(D_\mathrm{l}D_\mathrm{ls}/D_\mathrm{s})$
is chosen as a length scale, where $D_\mathrm{ls}$ is the
angular-diameter distance between lens and source. The derivative of
the lensing potential with respect to $x$ is the reduced deflection
angle at distance $x$,
\begin{equation}
  \alpha(x)=\frac{x}{|x|}\;.
\label{eq:sisalpha}
\end{equation}

For an NFW sphere, we choose $\xi_0=r_\mathrm{s}$ and define
$\kappa_\mathrm{s}\equiv\rho_\mathrm{s}r_\mathrm{s}\Sigma_\mathrm{cr}^{-1}$,
where $\Sigma_\mathrm{cr}=[c^2/(4\pi
G)]\,[D_\mathrm{s}/(D_\mathrm{l}D_\mathrm{ls})]$ is the critical
surface mass density for strong lensing. Its lensing potential then
reads
\begin{equation}
  \Psi(x)=4\kappa_{\rm s} g(x)\:,
\label{eq:nfwPsi}
\end{equation}
where
\begin{equation}
  g(x)=\frac{1}{2}\ln^2 \frac{x}{2}+\left\{
    \begin{array}{r@{\quad \quad}l}
      2\,\mbox{arctan}^2\sqrt{\frac{x-1}{x+1}} & (x>1) \\
     -2\,\mbox{arctanh}^2\sqrt{\frac{1-x}{1+x}} & (x<1) \\
      0 & (x=1)
    \end{array}\right.\;.
\end{equation}
This implies the deflection angle
\begin{equation}
  \alpha(x)=\frac{4\kappa_{\rm s}}{x} h(x) \:,
  \label{eq:nfwsalpha}
\end{equation}
with
\begin{equation}
  h(x)=\ln\frac{x}{2}+\left\{
    \begin{array}{r@{\quad \quad}l}
      \frac{2}{\sqrt{x^2-1}}\,
        \mbox{arctan}\sqrt{\frac{x-1}{x+1}} & (x>1) \\
      \frac{2}{\sqrt{1-x^2}}\,
        \mbox{arctanh}\sqrt{\frac{1-x}{1+x}} & (x<1) \\
      1 & (x=1)
    \end{array}\right.\;.
\end{equation}
\cite{meneghetti02,bartelmann96}.

\subsection{Elliptical model}

Lensing by galaxy clusters can only crudely be described by axially
symmetric models because their generally high level of asymmetry and
substructure changes their lensing properties qualitatively and
substantially. We thus perturb the axially symmetric lens models such
that their lensing potentials (\ref{eq:sisPsi})) and (\ref{eq:nfwPsi})
acquire elliptical isocontours. We define the ellipticity as
$e\equiv1-b/a$, where $a$ and $b$ are the major and minor axes of the
ellipse, respectively, and introduce it into (\ref{eq:nfwPsi}) by
substituting
\begin{equation}
  x\rightarrow\mathcal{X}=\sqrt{\frac{x_1^2}{(1-e)}+x_2^2(1-e)}\;,
\end{equation}
where $x_1$ and $x_2$ are the two Cartesian components of $x$,
$x^2=x_1^2+x_2^2$. The Cartesian components of the new deflection
angle are
\begin{eqnarray}
  \alpha_1 &=& \frac{\partial\psi}{\partial x_1}=
  \frac{x_1}{(1-e){\cal X}}\hat{\alpha}({\cal X})\;,\nonumber\\
  \alpha_2 &=& \frac{\partial\psi}{\partial x_2}=
  \frac{x_2(1-e)}{\cal X}\hat{\alpha}({\cal X})\;,
\label{eq:ellalpha}
\end{eqnarray}
where $\hat{\alpha}(\mathcal{X})$ is the unperturbed
(i.e.~axially-symmetric) deflection angle at the distance
$\mathcal{X}$ from the lens centre.

\subsection{Estimates}

Arcs form close to the tangential critical curve of a lens from
sources close to the tangential caustic. The size of the lensing cross
sections for the formation of tangential arcs is strictly connected to
the extent of the tangential critical curves and caustics: the longer
the critical curves and the corresponding caustics are, the larger are
the lensing cross sections. For axially symmetric lens models, the
tangential critical curve encloses a circle which encloses an average
convergence of unity, so that
\begin{equation}
  \frac{2}{\theta_\mathrm{t}^2}\,
  \int_0^{\theta_\mathrm{t}}\,\kappa(\theta')\,\theta'\d\theta'=1
\label{eq:meanKappa}
\end{equation}
defines the radius $\theta_\mathrm{t}$ of the tangential critical
curve. For singular isothermal spheres,
\begin{equation}
  \theta_\mathrm{t}=4\pi\,\left(\frac{\sigma_v}{c}\right)^2\,
  \frac{D_\mathrm{ls}}{D_\mathrm{s}}\;,
\label{eq:thetaSIS}
\end{equation}
which is approximately proportional to the lens mass. A similar closed
expression cannot be given for an NFW sphere. The tangential critical
radius of a set of concentrically superposed SIS lens models is simply
the sum of the tangential critical radii of the individual
components. This implies that adding a cD galaxy of
$\sim10^{13}\,h^{-1}\,M_\odot$ extends the tangential critical curve
of a cluster of $\sim5\times10^{14}\,h^{-1}\,M_\odot$ by a few per
cent only.

\begin{figure}\ifjournal\else[ht]\fi
  \includegraphics[width=\hsize]{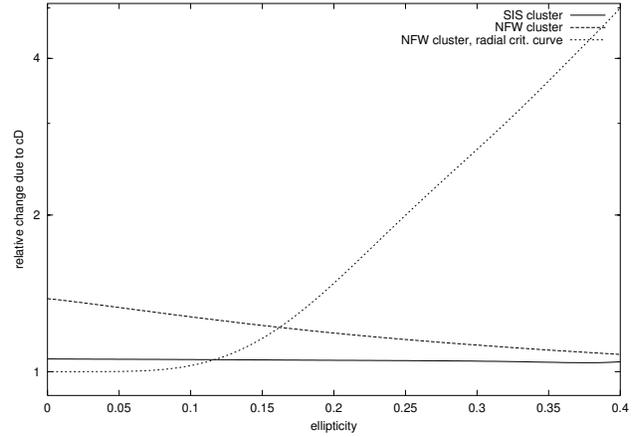}
\caption{The solid and the dashed curves show the ratios between the
  tangential critical radii of a cluster of virial mass
  $7.5\times10^{14}\,h^{-1}\,M_\odot$ with and without a cD galaxy of
  virial mass $10^{13}\,h^{-1}\,M_\odot$ added to its centre. They
  were obtained modelling the cluster with a SIS or NFW density
  profile, respectively. The dotted curve shows the ratios between the
  radial critical radii, obtained modelling the cluster with an NFW
  density profile. The ratios are given as functions of
  the ellipticity $e$ of the cluster potential, as described in the
  text. Since critical curves are not circular for $e>0$,
  we use their maximum cluster-centric distance instead.}
\label{fig:testE}
\end{figure}

However, the situation changes remarkably if the cD galaxy, modelled
as a SIS, is added to a cluster with an NFW density profile. The
comparatively flatter density profile near the cluster centre causes
the cD galaxy to have a much stronger effect, as Fig.~\ref{fig:testE}
illustrates.

The solid and dashed curves shown in the figure are the ratios between
the tangential critical radii of a cluster with virial mass
$7.5\times10^{14}\,h^{-1}\,M_\odot$ with and without a cD galaxy of
virial mass $10^{13}\,h^{-1}\,M_\odot$ added to its centre. They are
obtained for clusters with SIS and NFW density profiles,
respectively. The curves show how these ratios change with the
ellipticity $e$ of the cluster's lensing potential. The tangential
critical curves are circles only for $e=0$; for $e>0$, we use their
maximum cluster-centric distance instead.

Two results can be read off Fig.~\ref{fig:testE}. First, the effect of
a cD galaxy of fixed mass on the tangential critical curves of a
cluster of fixed mass is much more pronounced if the cluster has an
NFW rather than a SIS density profile; and second, the effect of the
cD decreases as the ellipticity of the cluster increases because the
increased shear of the cluster makes the additional surface mass
density relatively less important. For the axially symmetric NFW
cluster, the cD extends the tangential critical curve by as much as
$38\%$, although it has only a few per cent of the cluster's mass!
Numerical simulations suggest ellipticities of $\sim0.3$, for which
the effect drops $\sim12\%$.

Since we have good reasons to believe that clusters are better
described by NFW than by SIS density profiles, the figure illustrates
that an estimate of the effect of cD galaxies on strong lensing by
clusters based on SIS models alone can be highly misleading. In
addition, the change with cluster ellipticity is important because it
shows that the enhanced gravitational shear due to asymmetries and
substructure in the cluster potential reduces the impact of the cD
galaxy.

\begin{figure}\ifjournal\else[ht]\fi
  \includegraphics[width=\hsize]{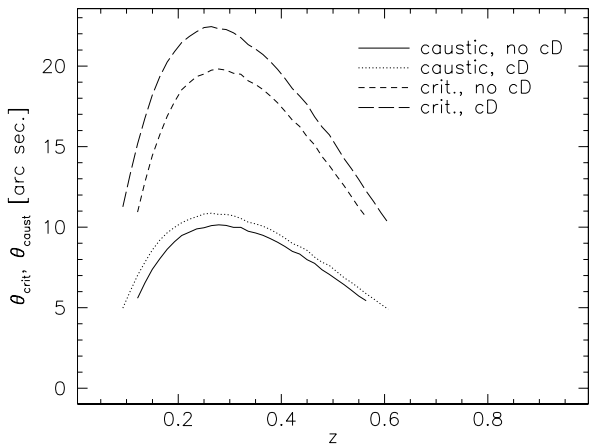}
\caption{The curves show, as functions of lens redshift, the maximum
  extent of tangential critical curves and corresponding caustics of a
  cluster with NFW density profile and virial mass
  $7.5\times10^{14}\,h^{-1}\,M_\odot$, with and without a cD galaxy
  with singular isothermal density profile and virial mass
  $10^{13}\,h^{-1}\,M_\odot$. An ellipticity of $e=0.3$ was assumed
  for the cluster's lensing potential, and the source redshift is
  $z_\mathrm{s}=1$. Obviously, the impact of a cD galaxy is highly
  insensitive to the lens redshift. The curves end where the maximum
  extent of the tangential critical curves drops below $10''$ because
  we are interested in the formation of large arcs only, which require
  a minimum size of the tangential critical curves.}
\label{fig:testZ}
\end{figure}

It is a second interesting question whether the impact of a cD galaxy
on a cluster lens changes with cluster redshift. Again, for two
concentrically superposed SIS lenses, the relative change in the
tangential critical radii is exactly independent of redshift, but this
does not need to hold true for cluster lenses with NFW density
profile. We show in Fig.~\ref{fig:testZ} as a function of lens
redshift the maximum extent of the tangential critical curve and the
corresponding caustic for a cluster with NFW density profile and
virial mass $7.5\times10^{14}\,h^{-1}\,M_\odot$, with and without a cD
galaxy with singular isothermal density profile and virial mass
$10^{13}\,h^{-1}\,M_\odot$. The cluster's lensing potential is assumed
to have ellipticity $e=0.3$, in good agreement with values found in
numerical simulations.

The curves end where the maximum extent of the tangential critical
curve drops below $10''$, because arcs formed near smaller tangential
critical curves can hardly be classified as large arcs. The curves
illustrate that the relative change in the dimensions of tangential
critical curves and caustics is remarkably insensitive to the lens
redshift, even if the cluster is assumed to have an NFW density
profile.

Finally, we consider the effect of a cD galaxy on a cluster's
efficiency for producing radial arcs. The dotted curve shown in
Fig.~\ref{fig:testE} is the ratio between the radial critical radii of
a cluster with and without a cD galaxy added to its center. The virial
masses of the cluster and the cD galaxy are the same as in the
previously discussed tests. Again, the cD galaxy is modelled as a SIS
while the cluster has an NFW density profile. As the figure shows, the
impact of the cD galaxy is negligible when the cluster is axially
symmetric. Indeed, the radial critical curve of such a cluster is a
very small circle around the cluster center. Adding a cD galaxy
modelled as a SIS does not change much because it dominates the
density profile only close to the cluster core. Since for a SIS the
derivative of the deflection angle, $\d\alpha/\d x$, is zero
everywhere, while the condition for having a radial critical curve in
an axially symmetric model is $\d\alpha/\d x=1$, the extent of the
radial critical curve is not changed by the central galaxy. For it to
be efficient, the radial critical curve or parts of it need to be
pushed outward where the density and deflection-angle profiles are no
longer dominated by the cD galaxy, and once this is achieved by the
ellipticity, the cD galaxy can introduce considerable change. We thus
expect that the impact of cD galaxies on the efficiency for producing
radial arcs of realistic cluster models might be much larger than for
tangential arcs.

We now turn to numerical cluster models in order to test the impact of
cD galaxies on cluster lensing with more realistic cluster mass
distributions. 

\section{Numerical models}
\label{section:nummod}

\subsection{Simulated Clusters}

We investigate the lensing properties of a sample of five numerically
simulated cluster-sized dark matter halos kindly made available by the
GIF collaboration \cite{kauffmann99}. The same clusters were used by
Bartelmann et al.~(1998). They were obtained from $N$-body simulations
performed in the framework of several cosmological models. In this
paper, we only use the simulations for a flat model with cosmological
constant ($\Lambda$CDM), which has $\Omega_0=0.3$,
$\Omega_{\Lambda}=0.7$ and $h=0.7$, and in an open model without
cosmological constant (OCDM, $\Omega_0=0.3$, $h=0.7$).

The initial matter density in these models is perturbed about the mean
according to a CDM power spectrum \cite{bond84} with primordial
spectral index $n=1$, normalised such that the local abundance of
massive galaxy clusters is reproduced (e.g. Viana \& Liddle 1996). The
complete list of cosmological parameters in these simulations is given
in Bartelmann et al.~(1998).

For each cosmological model, an initial simulation with $N=256^3$
particles in a box of $141\,h^{-1}\,\mathrm{Mpc}$ side length was
run. The mass of individual particles is $1.4\times10^{10}
\,h^{-1}\,M_\odot$. Clusters were obtained from initial cosmological
simulations as follows. High density regions were identified by means
of a standard friends-of-friends algorithm, selecting only the dense
cores of the collapsed objects. Around the centres of those, all
particles were collected which lie within a sphere of radius
$1.5\,h^{-1}\,\mathrm{Mpc}$, corresponding to the Abell radius. These
objects are taken as clusters. For our analysis, we took the five most
massive clusters in the simulation volumes.

As shown by Bartelmann et al.~(1998), the maximum efficiency for the
production of long and thin arcs is reached by these clusters when
their redshift is in the range $0.2\la z\la0.4$. Since our analytic
estimates showed that the lens redshift is highly irrelevant for the
impact of a cD on strong cluster lensing, we only take the simulation
snapshots at redshift $z_\mathrm{L}=0.275$, where the number of
simulated arcs is largest and thus the uncertainties in the
numerically determined cross sections are smallest.

\subsection{Lensing simulations}
\label{section:lenssim}

The lensing properties of the numerical clusters are studied using the
ray-tracing technique. Our method has been fully described in several
earlier publications (see Meneghetti et al.~2000, 2001). We thus give
only a brief description here, referring the reader to those papers
for further detail.

Each numerical cluster is centred within a cube of
$3\,h^{-1}\,\mathrm{Mpc}$ side length. Within this box, we compute the
three-dimensional density field $\rho$ by interpolating the mass
density on a regular grid of $256^3$ cells, using the \emph{Triangular
Shape Cloud} method (TSC; see Hockney \& Eastwood 1988). Finally, we
project $\rho$ along the three independent box sides, obtaining three
surface density maps for each cluster. These are used as lens planes.
Adopting this \emph{thin screen} approximation is justified because
the distances between the sources and the lens and between the lens
and the observer are much larger than the physical sizes of the galaxy
clusters.

We then propagate a bundle of $2048\times2048$ light rays through the
central quarter of each of these maps. We focus on the central parts
of the clusters because our analysis concerns their strong-lensing
properties only. Thus, our ray-tracing simulations resolve scales of
order $\sim0.2''$, at the lens redshift. The deflection angles of each
ray are computed by directly summing the contributions from each mass
element of the cluster.

Once the deflection angles are known, we reconstruct the images of a
large number of extended background sources. They are all placed on
the same plane, located at the redshift $z_\mathrm{s}=1$. Although
real sources are distributed in redshift, putting them at a single
redshift is justified for the present purposes because the lens
convergence changes very little with source redshift if the lens
redshift is substantially smaller, as is the case here.

The sources have elliptical shape, with axial ratios randomly drawn
from the range $[0.5,1]$, and their area is equal to that of a circle
of $1''$ diameter. We initially distribute them on a coarse regular
grid in the source plane. Then, their spatial density is iteratively
increased towards the caustic curves, where the lensing effects
strengthen. This increases the probability of producing long arcs and
thus the numerical efficiency of the method. In order to compensate
for this artificial source-density increase, we assign in the
following statistical analysis to each image a statistical weight
proportional to the area of the grid cell on which the source was
placed.

Image reconstruction and classification is done following the
technique introduced by Bartelmann \& Weiss (1994) and used in
Meneghetti et al.~(2000, 2001, 2003). For each image, we measure its
length and width. This results in a catalogue of simulated images
which is subsequently analysed statistically.

\subsection{Inclusion of the cD galaxy}

Being linear function of mass, the total deflection angle of a ray
passing through a mass distribution is the sum of the contributions
from each mass element of the deflector. Therefore, in the case of a
galaxy cluster, we can decompose the cluster lens into its smoothed
dark matter component, plus the granular component contributed by its
galaxy population (see also Meneghetti et al.~2000). For both the
cluster and the galaxies, the main constituent is given by the dark
matter which forms their halos. Our model of the cluster containing a
cD galaxy can thus be fairly simple; we take the smoothed dark matter
distribution obtained from the numerical simulations described above,
and introduce a dark-matter halo resembling the galaxy. For each ray
traced through the lens plane, we compute the deflection angle by
summing the contributions from the cluster itself and the galaxy
haloes.

Using this approach, the reduced deflection angle of each ray
parameterised by its cell numbers $(i,j)$ on the grid in the image
plane is given by
\begin{equation}
  \vec\alpha_{ij}=\vec\alpha_{ij}^\mathrm{cl}+
  \vec\alpha_{ij}^\mathrm{cD}\;,
\end{equation} 
where $\vec\alpha_{ij}^\mathrm{cl}$ and $\vec\alpha_{ij}^\mathrm{cD}$
are contributions to the deflection angle from the original cluster
and the cD galaxy, respectively.

The values of $\vec\alpha_{ij}^\mathrm{cD}$ depend on the model for
the cD galaxy. We first apply two axially symmetric models, namely
spheres with the NFW or the singular isothermal density
profile. Second, we also apply the pseudo-elliptical NFW lens model
\cite{meneghetti02} in order to account for the possible elongation of
the matter distribution of the cD galaxy.

cD galaxies typically appear to be of elliptical shape, with isophotal
axis ratios $b/a\sim0.8$ \cite{porter91}. Moreover, the orientation of
the brightest cluster ellipticals is usually not random, but
correlates well with that of their host cluster
\cite{sastry68,rood72,austin74,carter80,binggeli82,struble85,rhee87,lambas88,garijo97}. Asymmetries
in the lensing matter distribution are known to improve the ability of
the cluster to produce long and thin arcs. We also expect that the
impact of a cD galaxy described by an elliptical model is largest when
its orientation is aligned with the elongation direction of the host
galaxy cluster. In the case of different orientations, the cD galaxy
tends to circularise the mass distribution of the cluster in its
central region. In order to quantify this effect, we carry out two
sets of simulations; in the first, we randomly choose the orientation
of the cD galaxy inside the cluster, while in the second the
orientation is chosen such that the directions of the major and minor
axes of the galaxy align with the major and minor eigenvalues of the
cluster's deflection angle field, respectively. The galaxy ellipticity
is assumed to be $e=0.2$ \cite{porter91}.

The position of the cD galaxy is chosen to coincide with the minimum
of the deflection angle field of the numerical galaxy cluster. This
choice intentionally maximises the contribution of the cD galaxy to
the total deflection angles.
 
\section{Results}
\label{section:results}

\begin{figure*}\ifjournal\else[ht]\fi
\centering
  \includegraphics[width=.25\hsize]{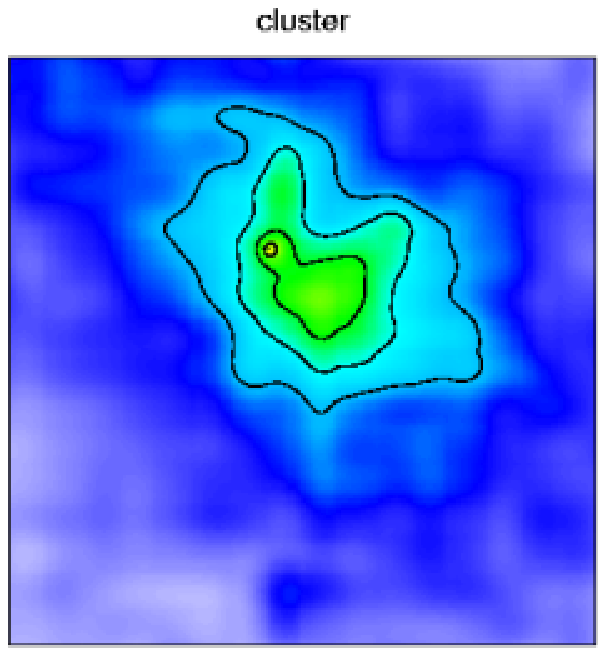}
  \includegraphics[width=.25\hsize]{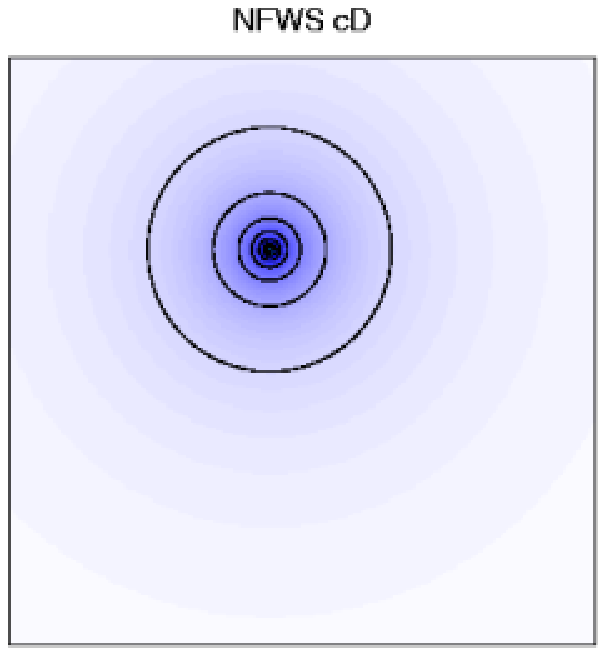}
  \includegraphics[width=.25\hsize]{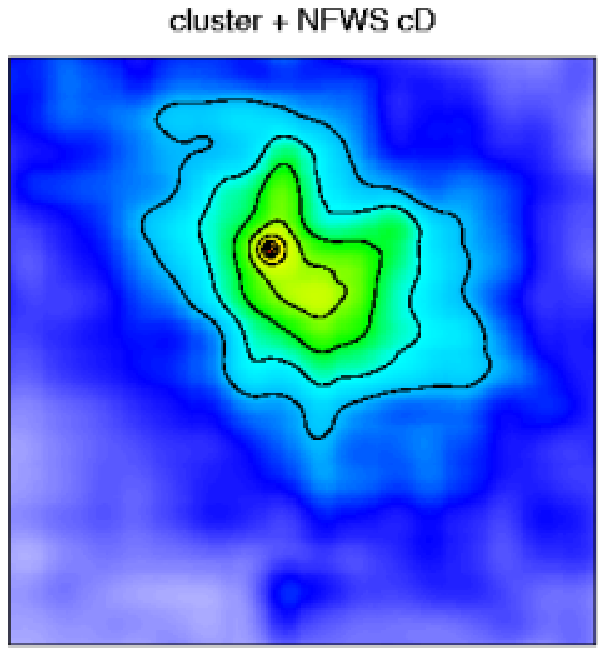}
  \includegraphics[width=.25\hsize]{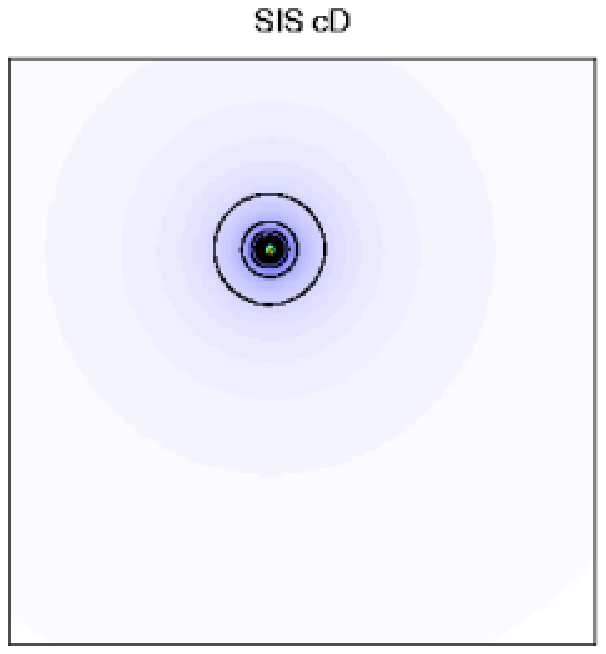}
  \includegraphics[width=.25\hsize]{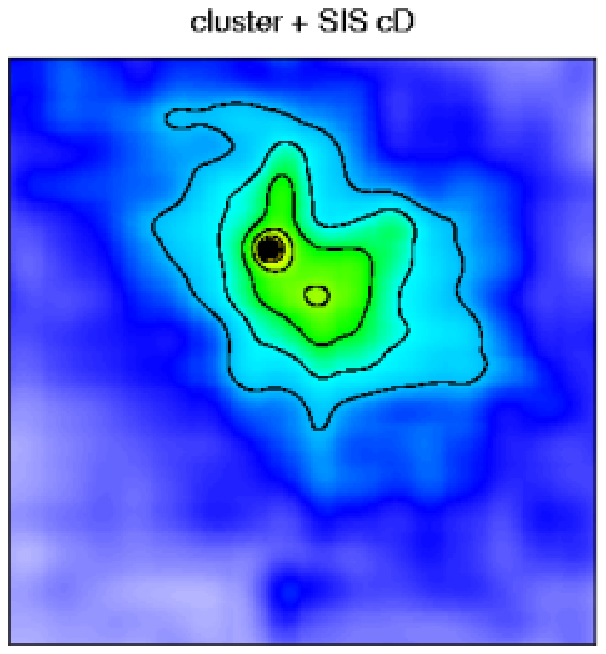}
  \includegraphics[width=.25\hsize]{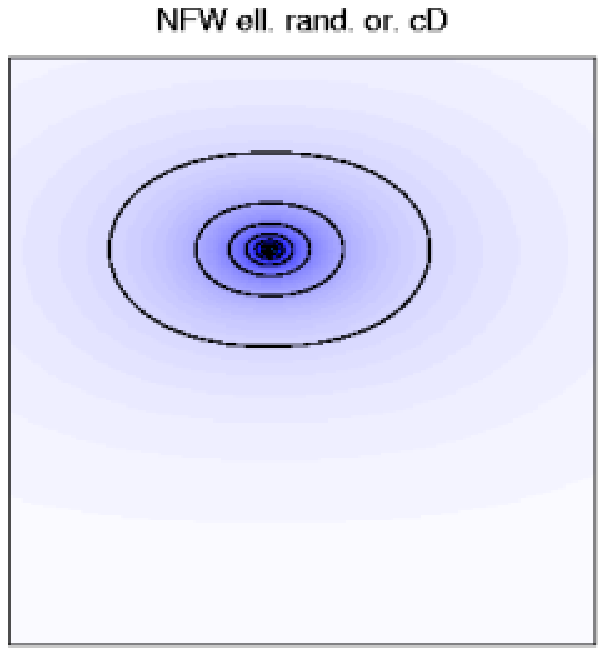}
  \includegraphics[width=.25\hsize]{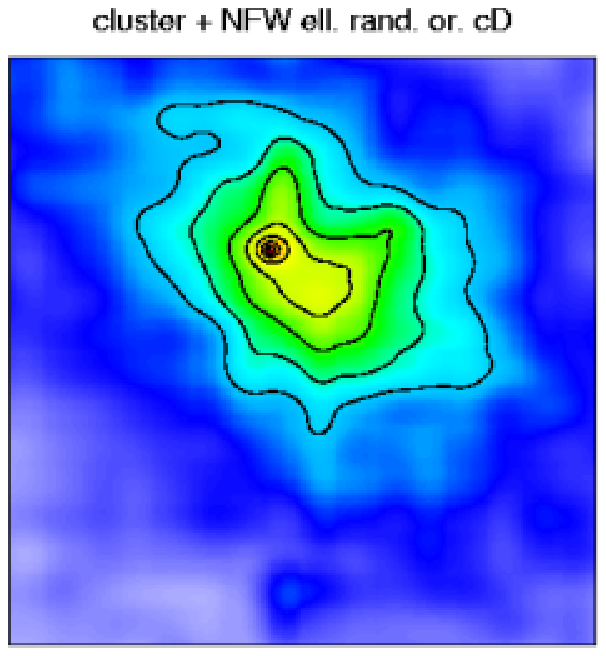}
  \includegraphics[width=.25\hsize]{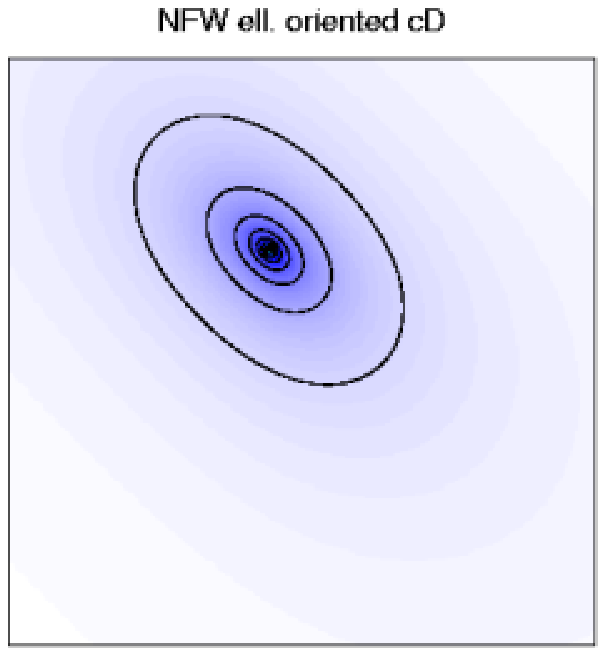}
  \includegraphics[width=.25\hsize]{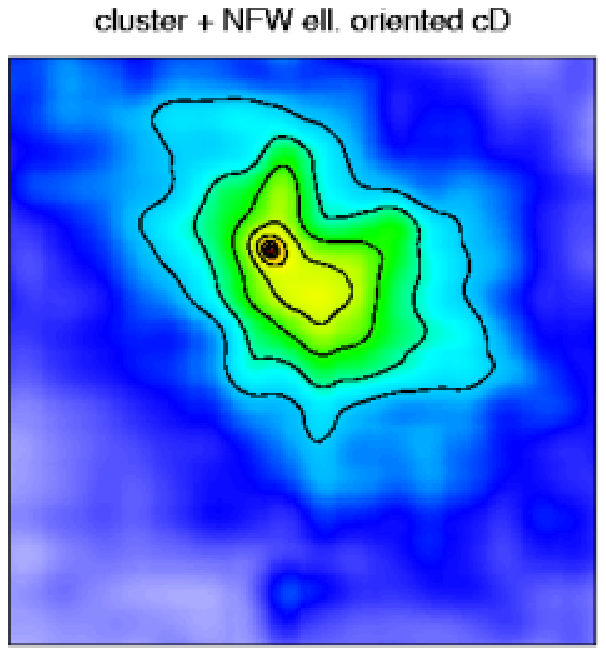}
  \includegraphics[width=.32\hsize]{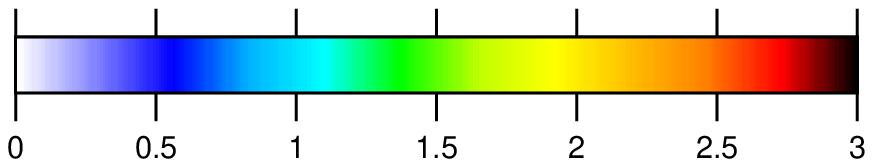}
\caption{Examples of convergence maps for one of the clusters in the
  GIF sample and for the cD galaxy models. Starting from the top-left
  panel: (a) cluster without the cD galaxy; (b) cD galaxy of mass
  $M_\mathrm{cD}=5\times10^{13}\,h^{-1}\,M_\odot$ with NFW density
  profile; (c) cluster with the cD galaxy modelled as in (b); (d) cD
  galaxy with the same mass, but with SIS density profile; (e) cluster
  with the cD galaxy modelled as in (d); (f) pseudo-elliptical NFW
  model randomly oriented with respect to the cluster mass
  distribution; (g) cluster with the cD galaxy modelled as in (f); (h)
  pseudo-elliptical NFW model with its orientation aligned with that
  of the cluster mass distribution; (i) cluster with the cD galaxy
  modelled as in (h). The scale of each panel is $\sim 200\,h^{-1}$kpc
  or $\sim 70$ arcsec.}
\label{figure:convergencemaps}
\end{figure*}

\subsection{Convergence maps and critical curves}

We describe in this Section how the cD galaxy changes the projected
mass distribution of the numerical clusters. The surface density map
of a lens can be reconstructed from the deflection angles once the
effect of the cD galaxy has been included. Indeed, the scaled surface
density $\kappa$ (or convergence) at each position $\vec{x}$ on the
lens plane is given by
\begin{equation} 
  \kappa(\vec{x})=\frac{\Sigma(\vec{x})}{\Sigma_\mathrm{cr}}=
  \frac{1}{2}\vec{\nabla}\cdot\vec{\alpha}(\vec{x})\;.
\label{equation:convergence}
\end{equation} 

\begin{figure*}\ifjournal\else[ht]\fi
  \centering
  \includegraphics[width=0.75\hsize]{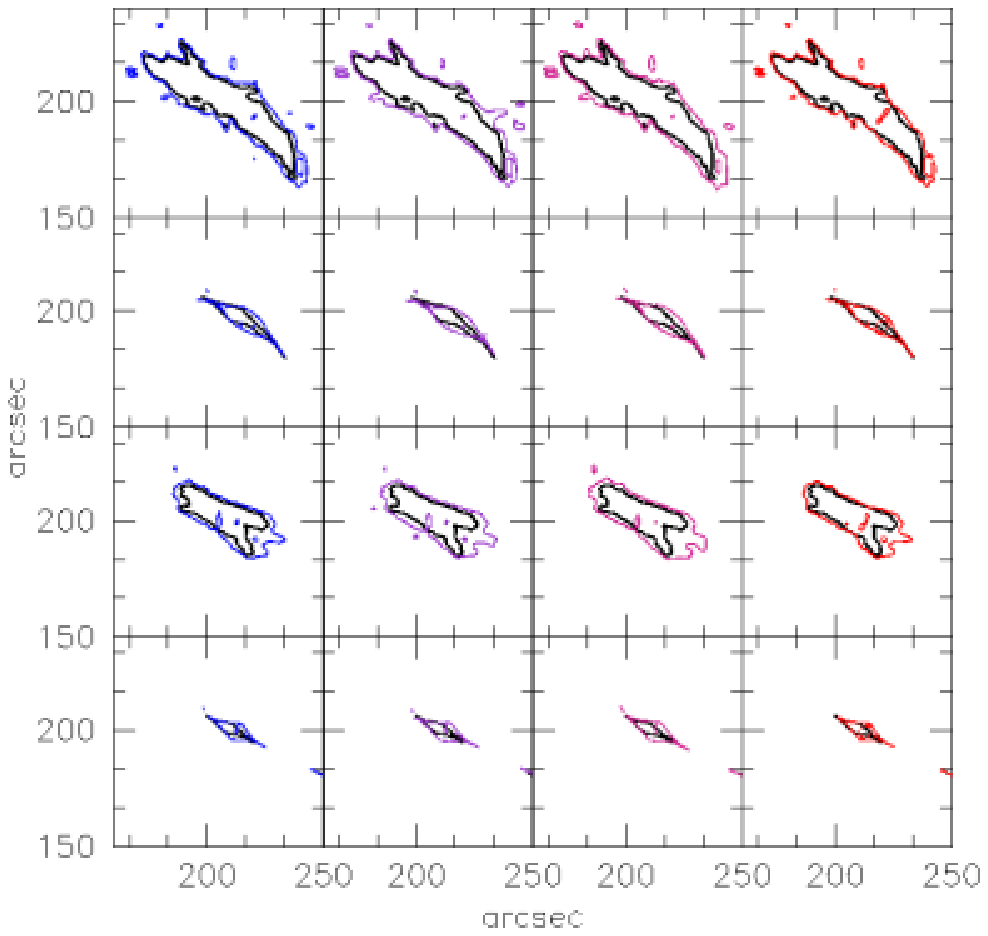}
\caption{Examples of critical curves and caustics. The first row of
  panels shows how the critical curves for one projection of the most
  massive cluster ($M\sim 10^{15}\,h^{-1}\,M_{\odot}$) in the GIF
  sample change in response to the inclusion of the cD galaxy. In each
  panel, we draw the original critical curves (bold curves) and those
  modified by the cD (thin curves). We show the results obtained by
  modelling the cD galaxy as an NFW sphere (first column from the
  left); using the pseudo-elliptical model randomly oriented (second
  column) and aligned with the cluster mass distribution (third
  column), and as a SIS (fourth column). The panels in the second row
  show the corresponding caustic curves. Similarly, we plot in the
  third and in the fourth row of panels the critical curves and the
  caustics for one projection of the least massive cluster in the
  sample ($M\sim 3\times 10^{14}\,h^{-1}\,M_{\odot}$). The mass of the
  cD galaxy is $M_\mathrm{cD}=5\times10^{13}\,h^{-1}\,M_\odot$, and
  the background cosmology is the $\Lambda$CDM model. The physical
  scale of each panel is $\sim 265\,h^{-1}$kpc.}
\label{figure:critcurves}
\end{figure*}

Some examples for convergence maps are shown in
Fig.~(\ref{figure:convergencemaps}). The original convergence map of
the cluster, i.e.~before adding the cD galaxy, is plotted in the top
left panel. Then, starting from the top central panel, we show the
convergence maps for all cD galaxy models we use, followed by the
convergence map of the cluster including the cD galaxy: the NFW sphere
(top central and top right panels), SIS (middle left and middle
central panels), the pseudo-elliptical NFW model both randomly
oriented (middle right and bottom left panels) and aligned with the
orientation of the cluster mass distribution (bottom central and
bottom left panels). The virial mass of the cD galaxy is the largest
we have tested in our simulations
($5\times10^{13}\,h^{-1}\,M_\odot$). The angular scale of each plot is
approximately $\sim50''$.

The contours of the convergence maps show that the presence of the cD
galaxy affects the matter distribution of the cluster only in its very
central part. The size of the region where the convergence enhancement
is significant depends on the model which is used to describe the cD
galaxy. For example, the SIS has a steeper density profile at the
centre and thus its contribution to the host cluster's surface density
is significant only in a very limited region. On the other hand, due
to its shallower density profile, the mass of a cD galaxy modelled as
an NFW sphere or as a pseudo-elliptical NFW model is distributed
across a larger region around the cluster centre. Moreover, when the
pseudo-elliptical model is used, convergence contours appear to be
more extended and particularly elongated if the cD galaxy is aligned
with the orientation of the underlying cluster mass distribution.

Arcs form along the critical curves, where the determinant of the
Jacobian matrix $A(\vec{x})=\partial{\vec{y}}/\partial{\vec{x}}$
vanishes. Given that the dimension-less coordinates on the source
plane $\vec{y}$ and on the lens plane $\vec{x}$ are related by the
lens equation, $\vec{y}=\vec{x}-\vec{\alpha}(\vec{x})$, the position
of the critical curves can be easily found through numerical
derivatives of the deflection angles. Then, using the lens equation,
we can find the position of the caustics, near which sources must be
located in order to be strongly magnified.

We show in Fig.~(\ref{figure:critcurves}) some examples of critical
curves and caustics for the most massive (first two rows of panels)
and the least massive (last two rows of panels) clusters in our sample
(in the $\Lambda$CDM model). In each panel, we plot the results before
adding the cD galaxy (thick curves), and after including the
contribution to the deflection angles from a cD galaxy with a virial
mass of $5\times10^{13}\,h^{-1}\,M_\odot$ (thin curves). In the first
column on the left, we show the results obtained by modelling the cD
as an NFW sphere; in the second column, we plot the critical curves
and caustics found modelling the galaxy as pseudo-elliptical, randomly
oriented NFW models; in the third column, we show the results for the
pseudo-elliptical NFW model aligned with the orientation of the
underlying cluster matter distribution; and in the fourth column we
finally show the critical curves and the caustics obtained modelling
the cD as a SIS.

As expected, the critical curves and the caustics appear to be pushed
towards larger cluster-centric radii when the cD galaxy is included
into the simulations. The largest effect is produced by the
pseudo-elliptical cD if its orientation is aligned with that of the
host cluster. The singular isothermal cD has the least effect on the
critical curves. Of course, the impact of the cD galaxy is different
on clusters with different masses. For example, the caustics of the
most massive lens demonstrate that they are affected by the presence
of the cD only far from the cusps. This is not the case for the least
massive cluster, where also the cusp positions are
pushed away from the lens centre. Also, the critical curves appear to
be more extended once the cD is added to the least massive cluster. Of
course, this is due to the fact that, by adding the cD, we are
changing the cluster mass in its central region by a different
percentage. Similar results are found for clusters in the OCDM model.
    
\subsection{Tangential arcs}

\begin{figure*}\ifjournal\else[ht]\fi
  \centering
  \includegraphics[width=.25\hsize]{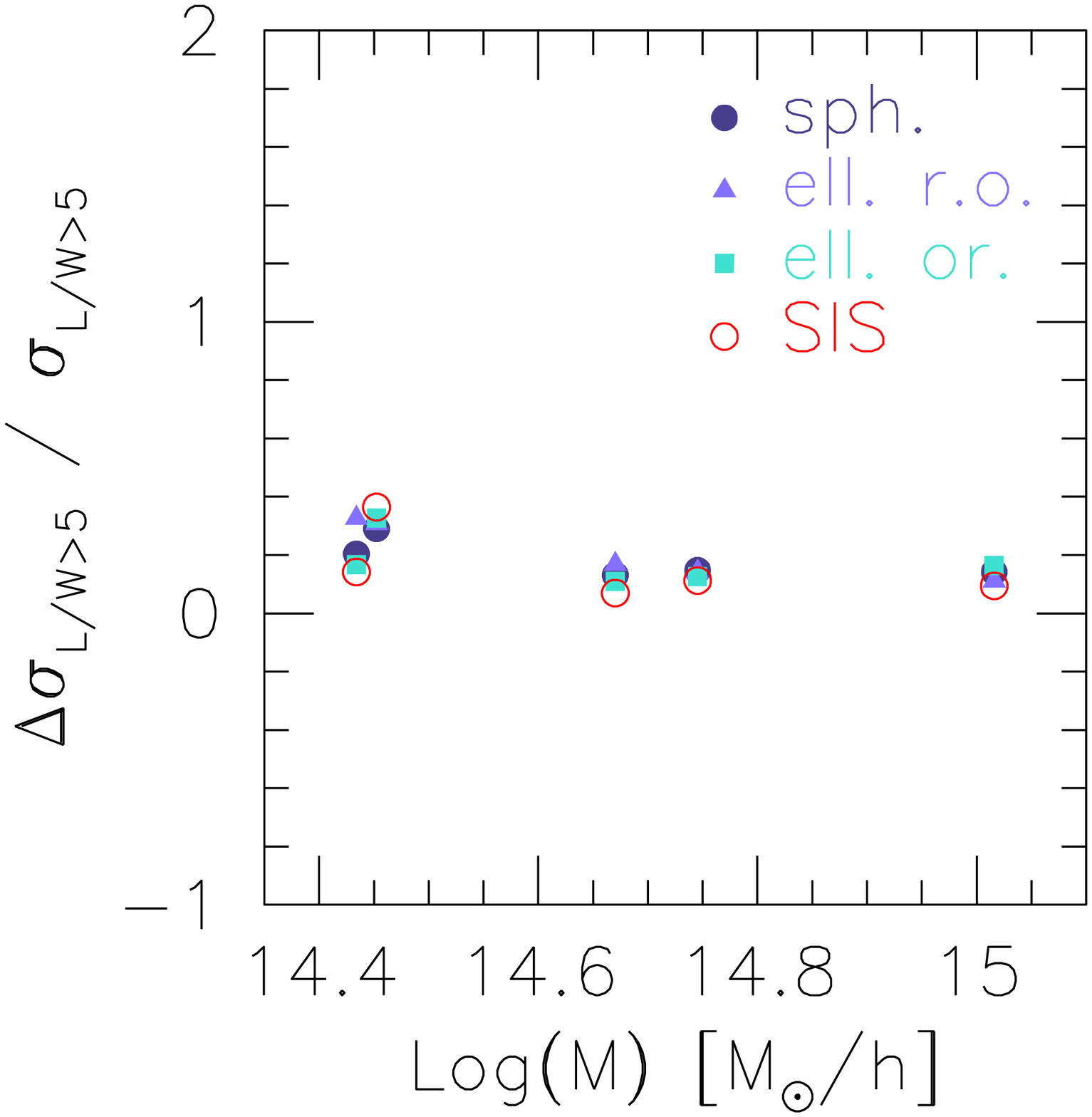}
  \includegraphics[width=.25\hsize]{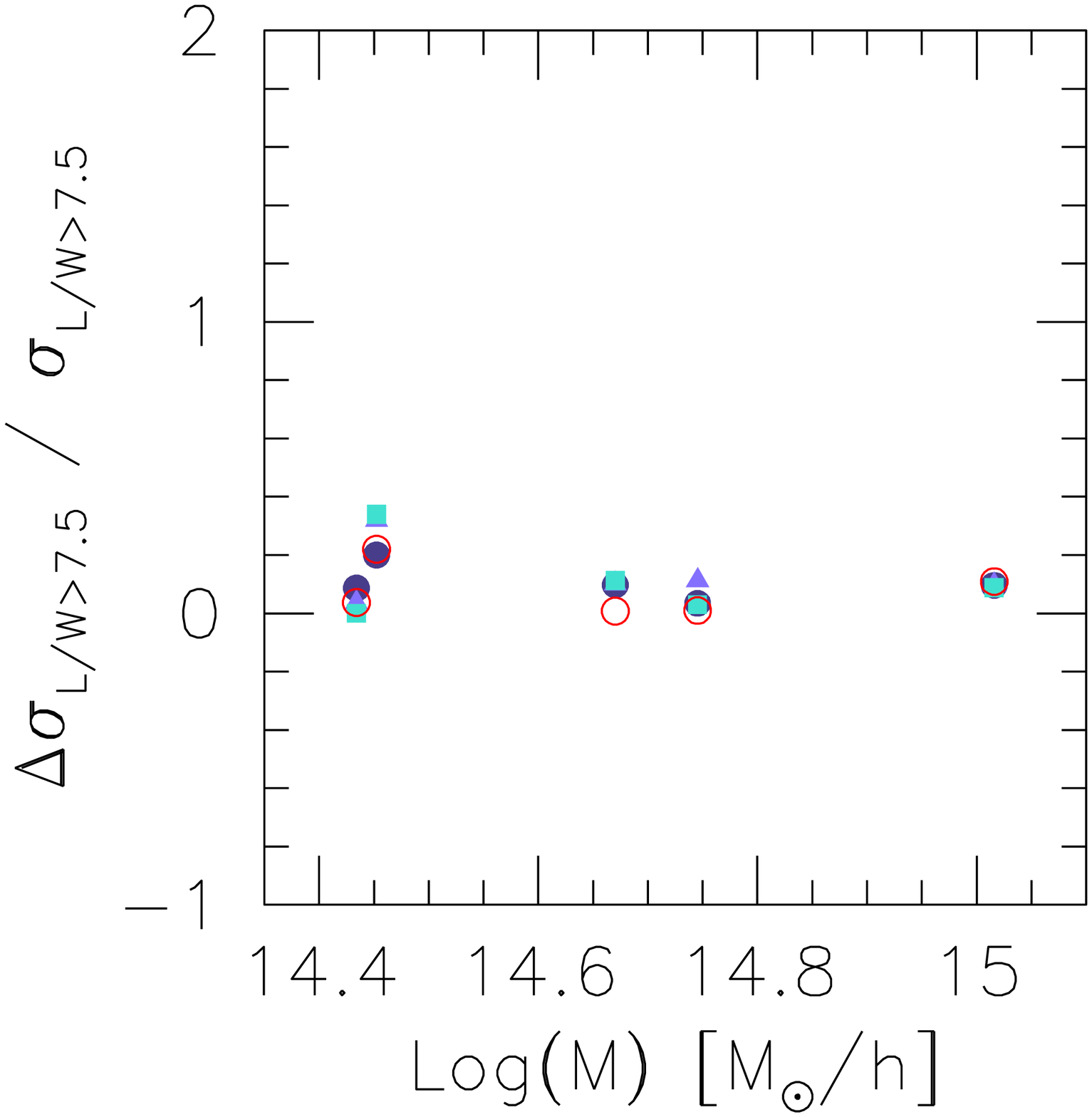}
  \includegraphics[width=.25\hsize]{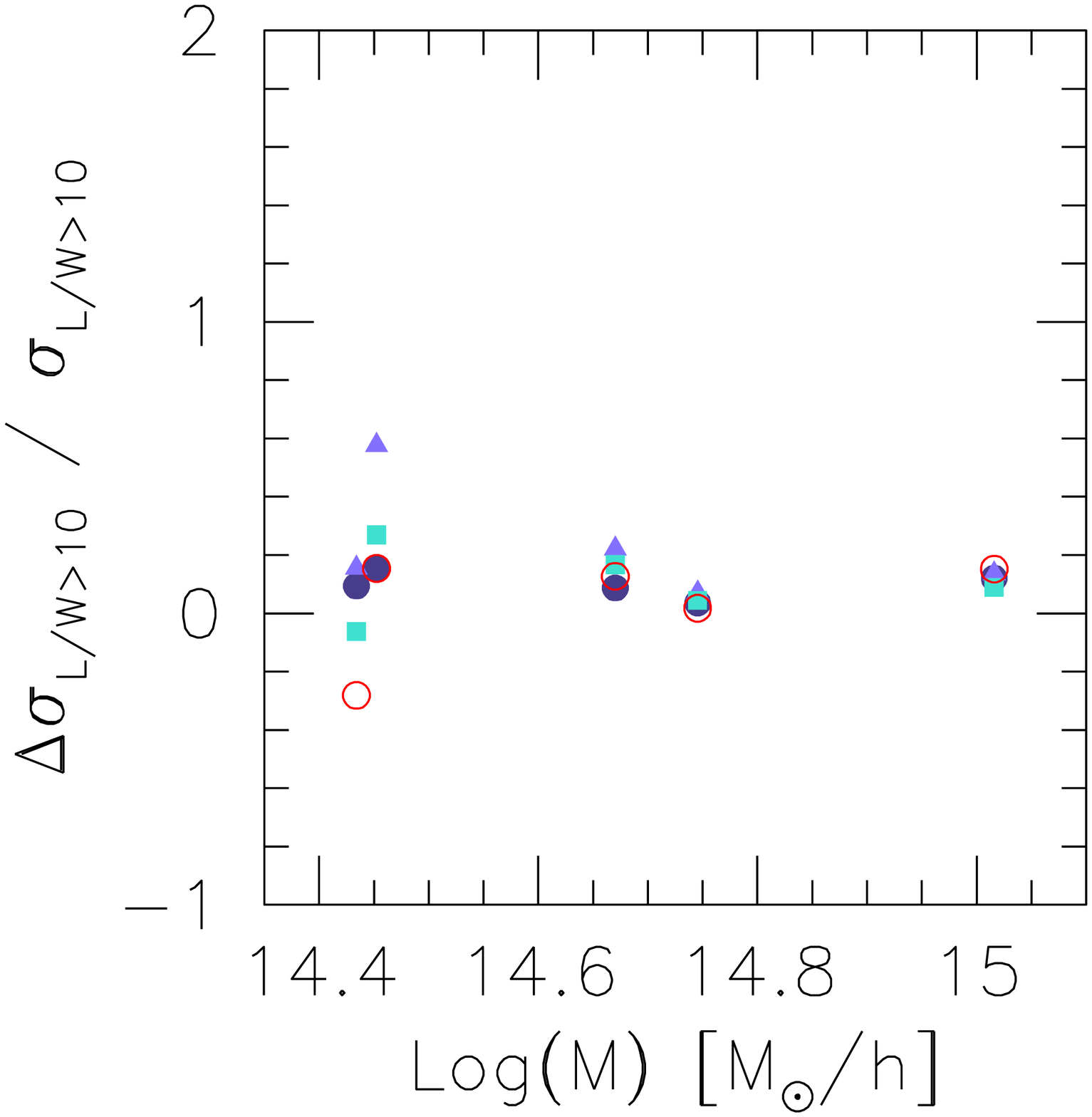}
  \includegraphics[width=.25\hsize]{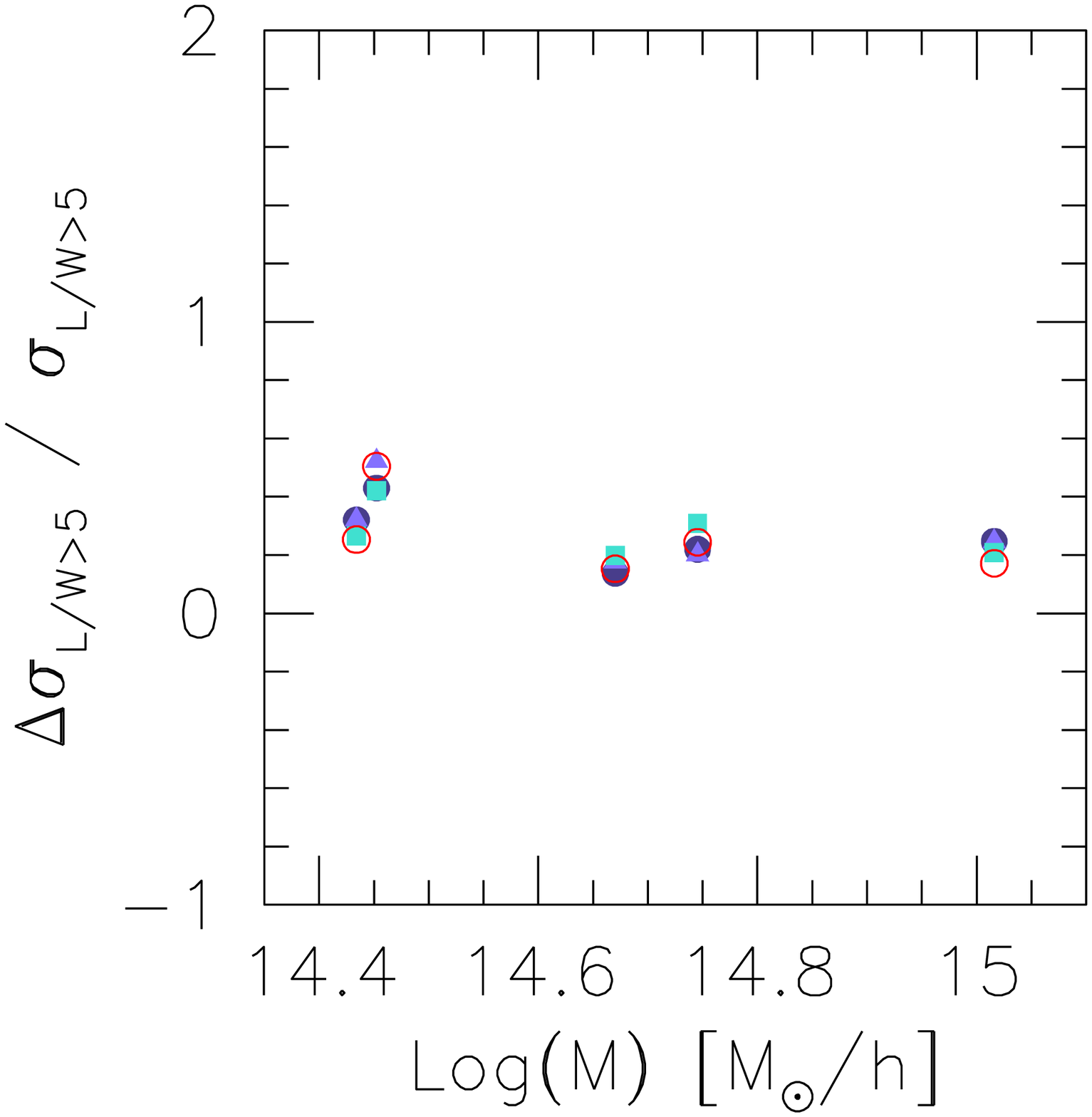}
  \includegraphics[width=.25\hsize]{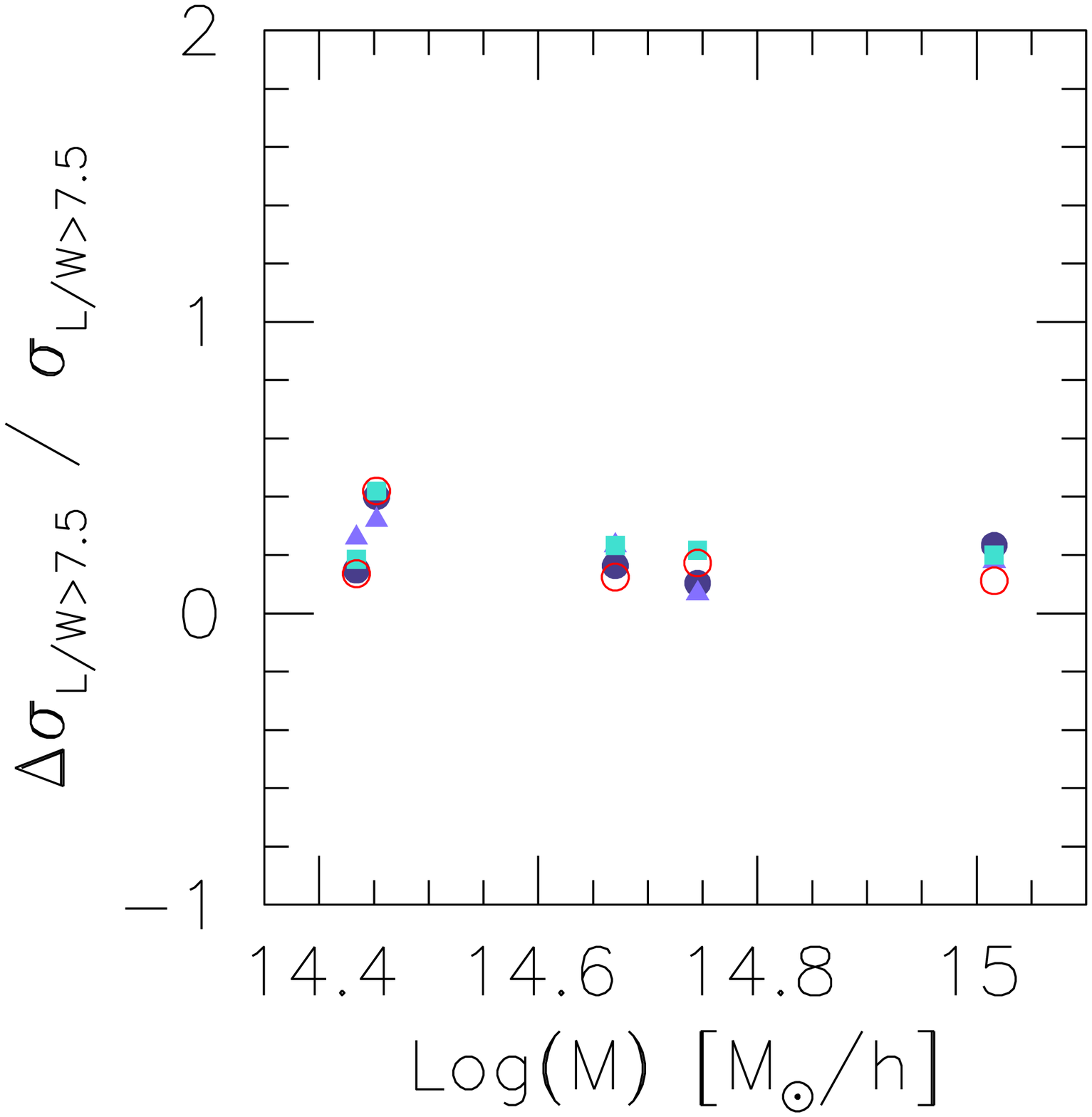}
  \includegraphics[width=.25\hsize]{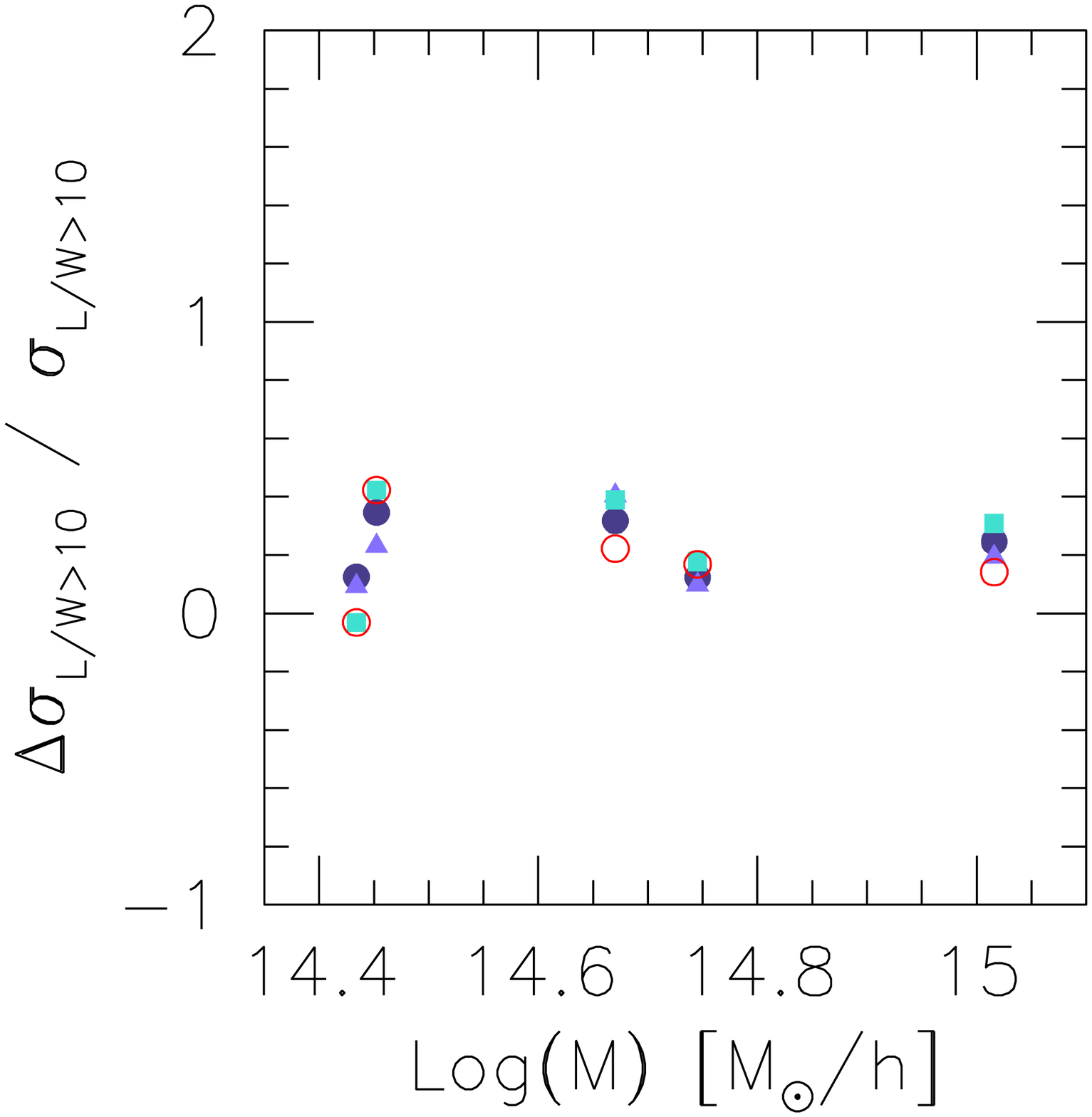}
  \includegraphics[width=.25\hsize]{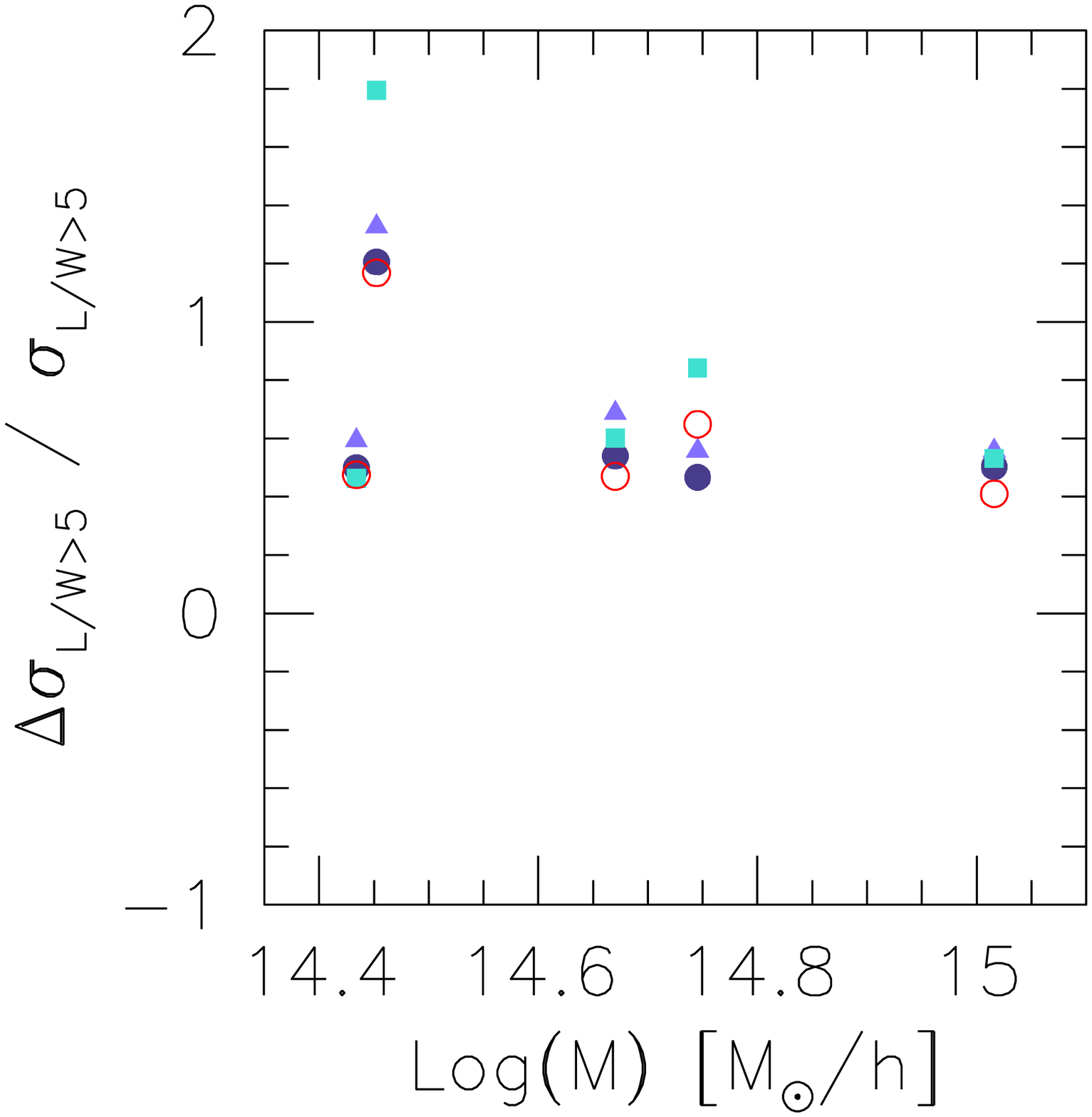}
  \includegraphics[width=.25\hsize]{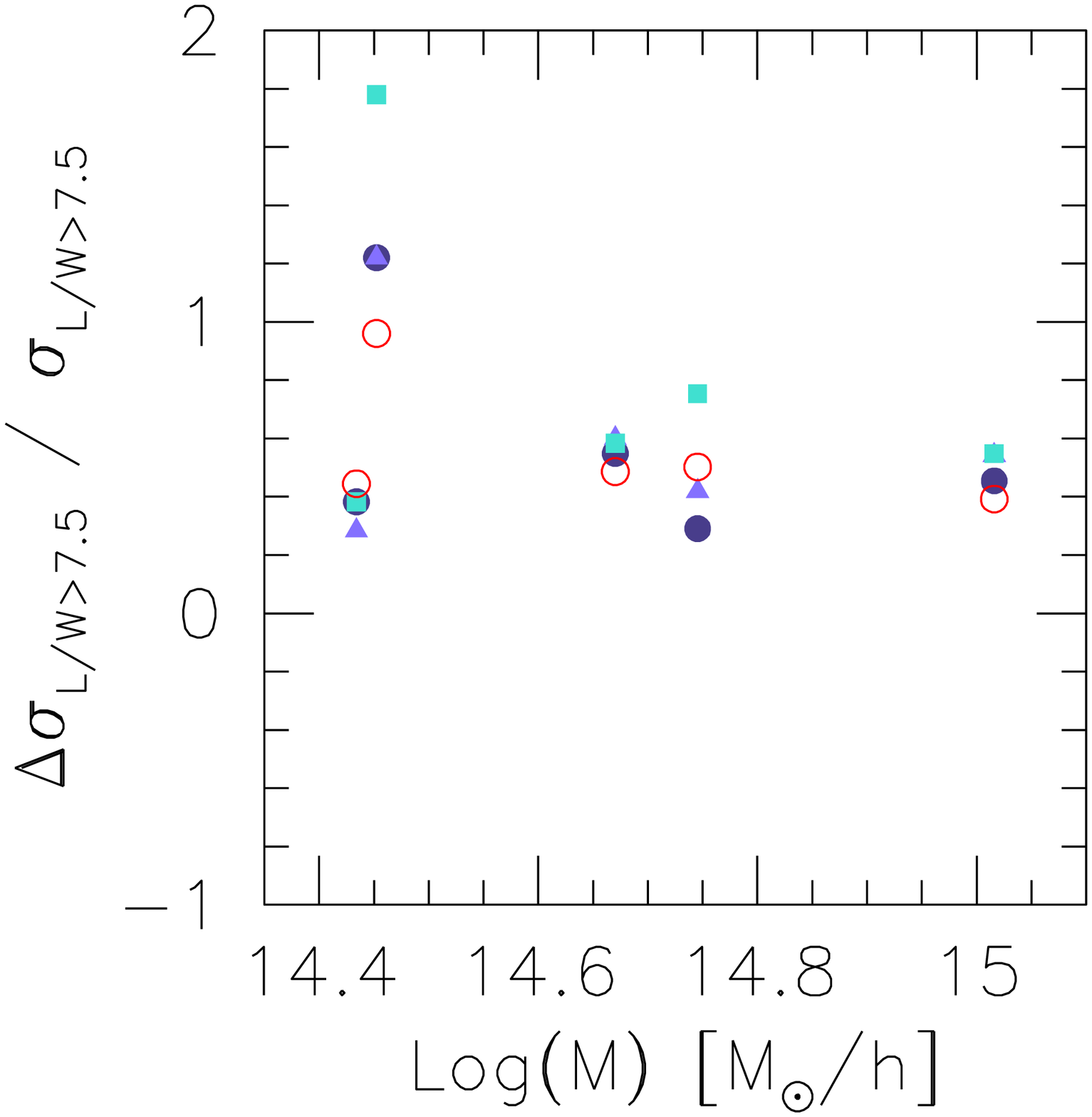}
  \includegraphics[width=.25\hsize]{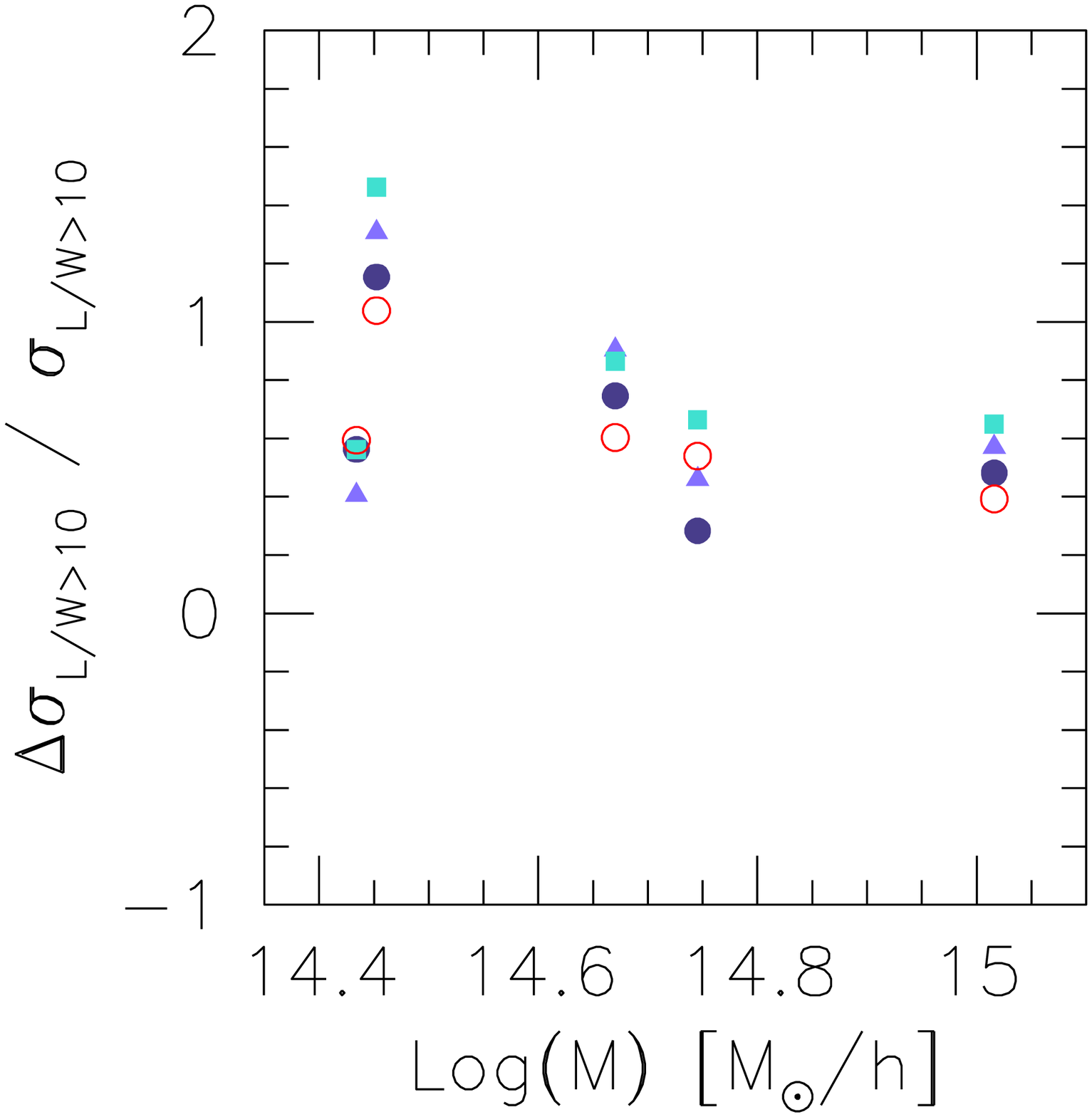}
\caption{Relative change in the cross sections for arcs with
  length-to-width ratios exceeding $5$ (left column), $7.5$ (central
  column) and $10$ (right column) as a function of cluster mass for
  the numerically simulated clusters in the $\Lambda$CDM
  model. Results are shown for three different masses of the cD
  galaxy: $5\times10^{12}\,h^{-1}\,M_\odot$ (top panels),
  $10^{13}\,h^{-1}\,M_\odot$ (middle panels) and
  $5\times10^{13}\,h^{-1}\,M_\odot$ (bottom panels). Filled circles,
  triangles and squares mark the results obtained modelling the cD as
  an NFW sphere, a pseudo-elliptical NFW model with random
  orientation, and aligned with the orientation of the host cluster,
  respectively; open circles show the results found modelling the cD
  galaxy as a SIS.}
\label{figure:crosssec_lambda}
\end{figure*}

\begin{figure*}\ifjournal\else[ht]\fi
  \centering
  \includegraphics[width=.25\hsize]{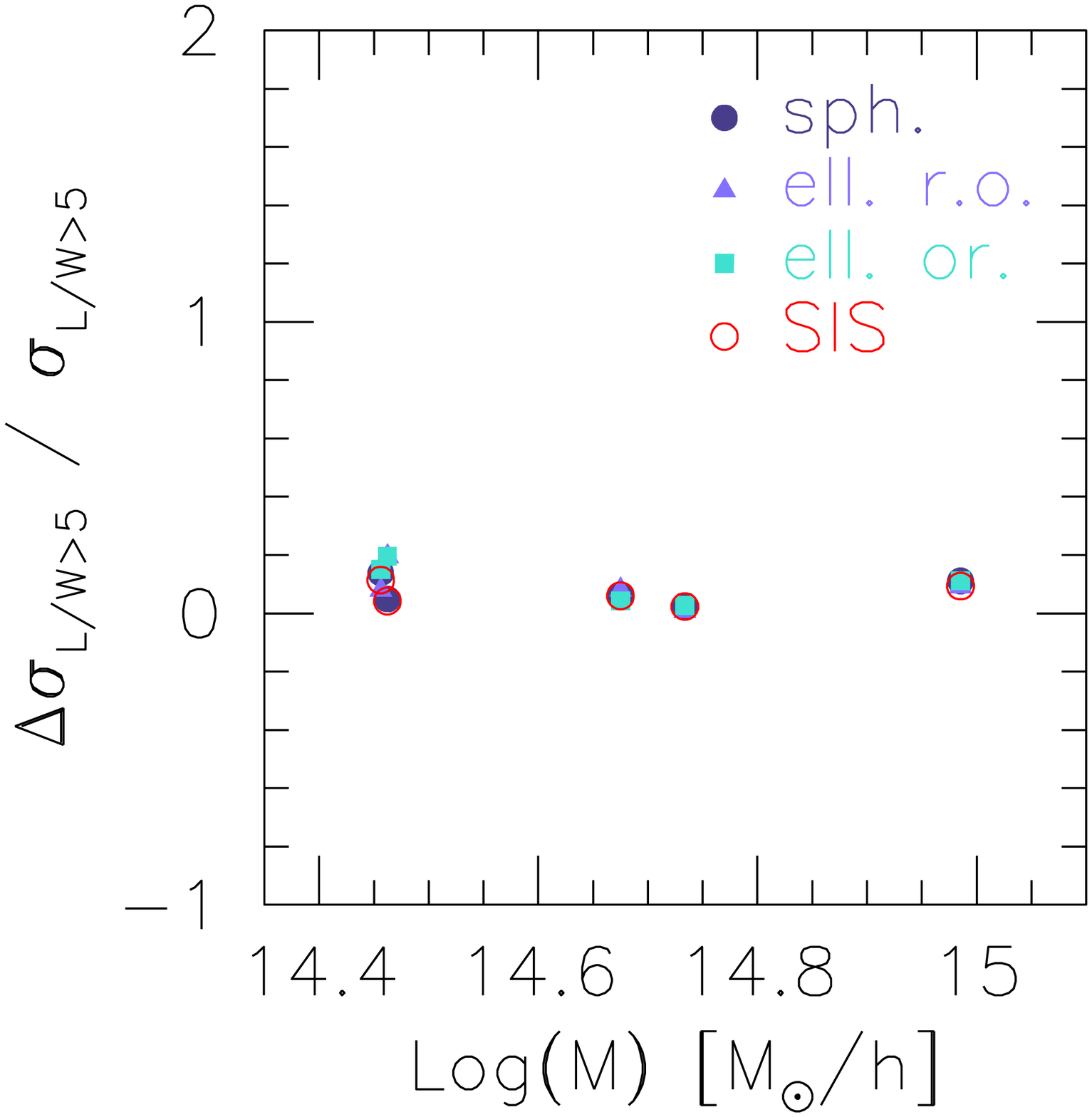}
  \includegraphics[width=.25\hsize]{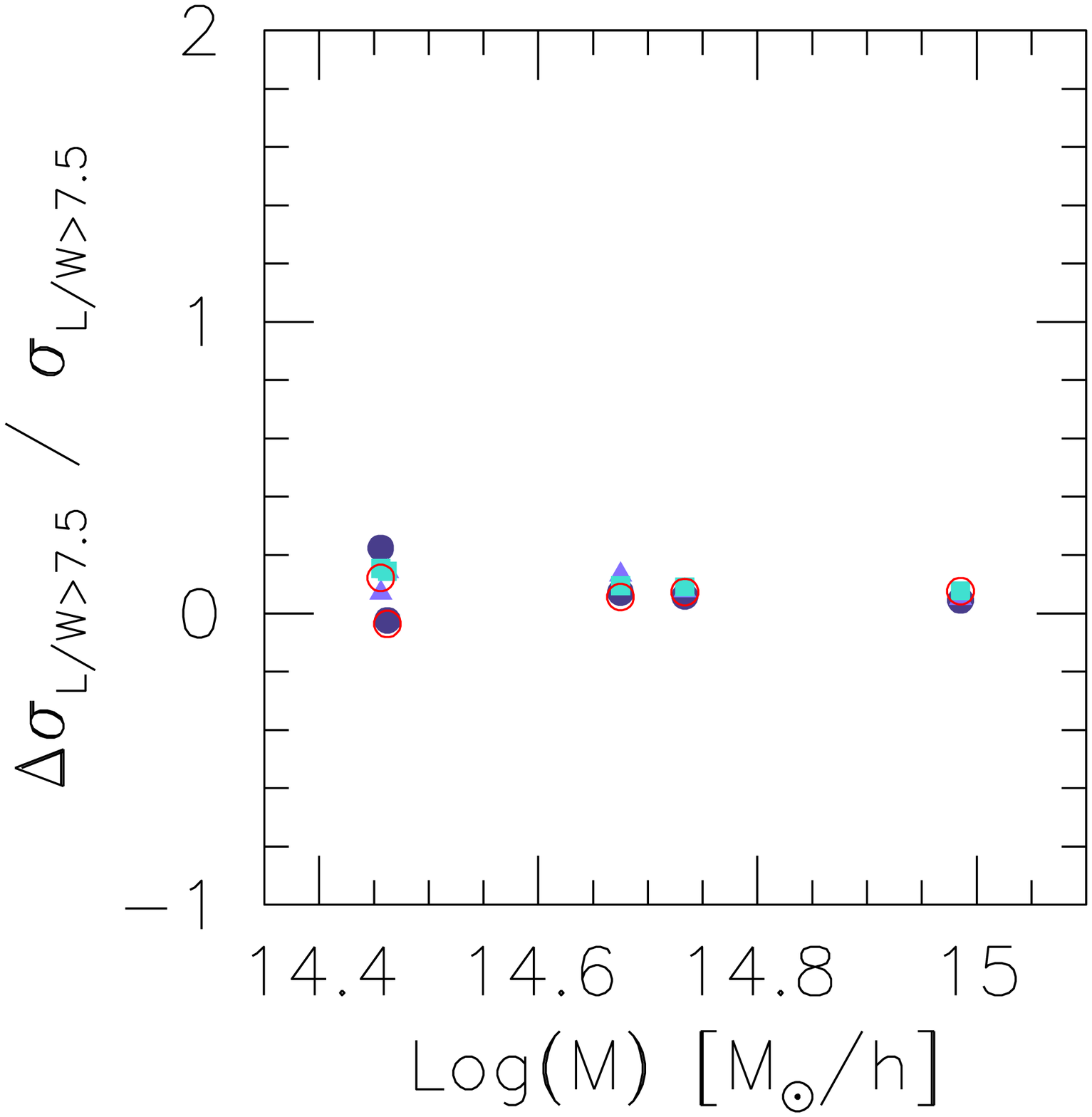}
  \includegraphics[width=.25\hsize]{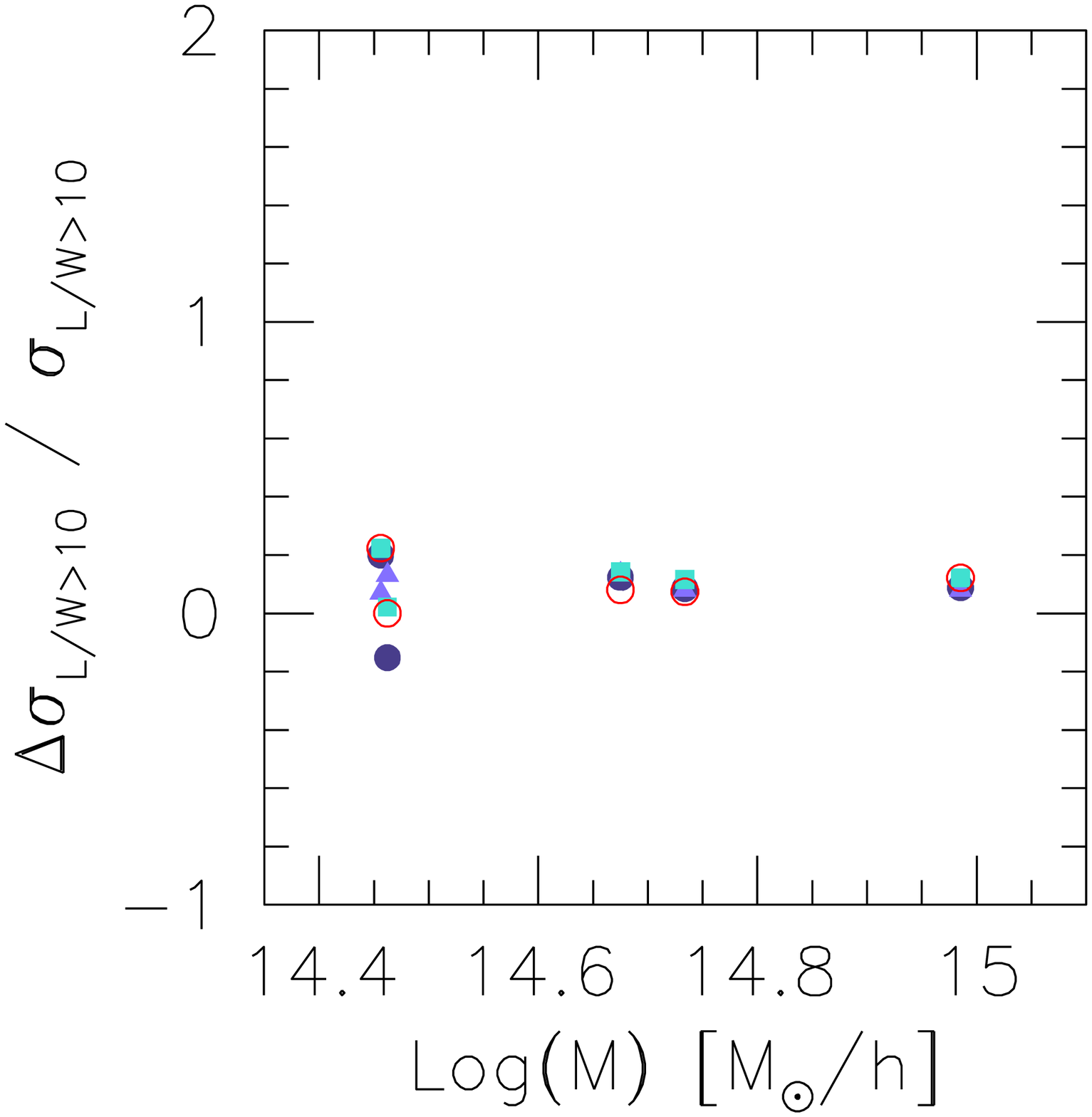}
  \includegraphics[width=.25\hsize]{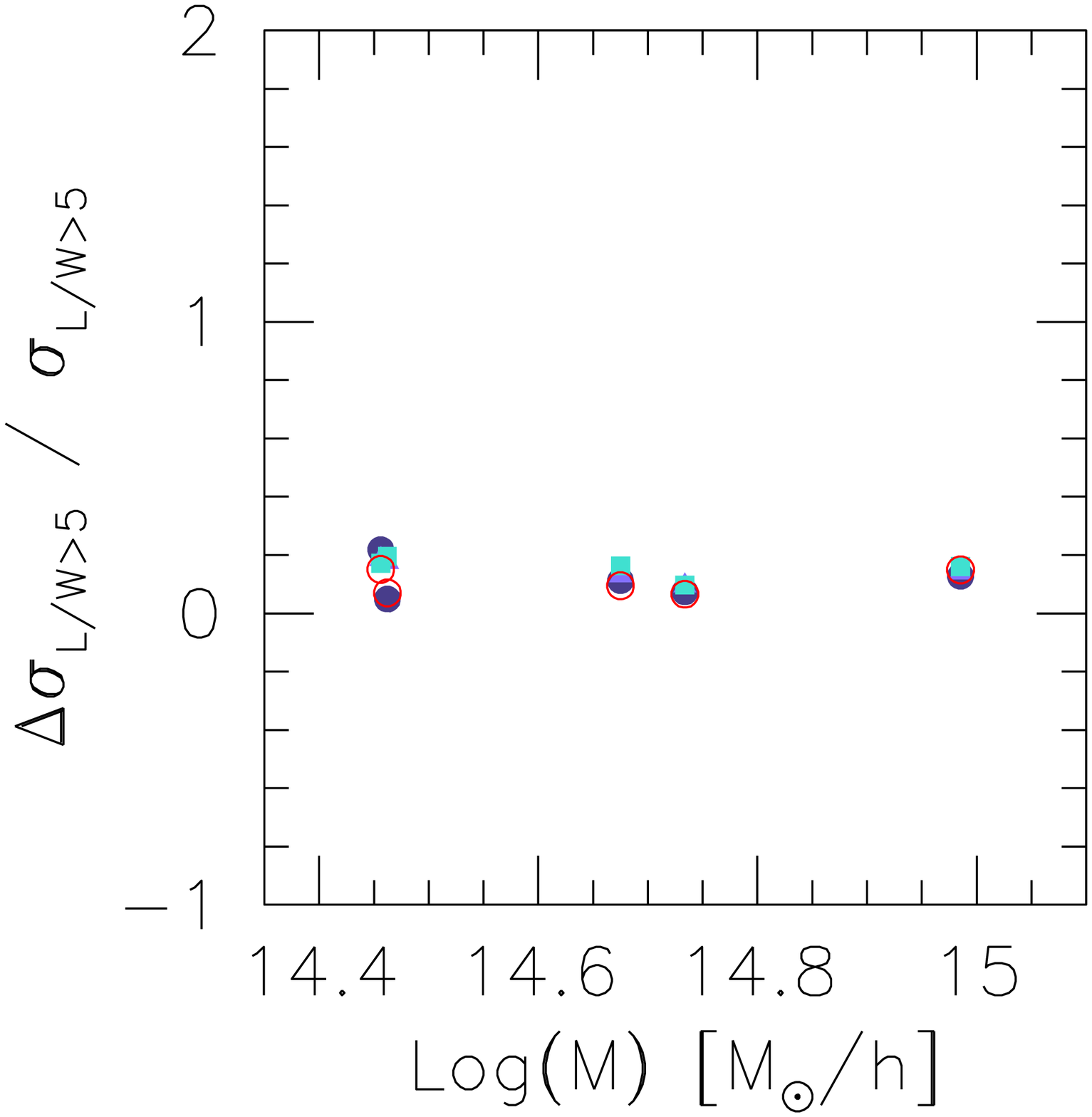}
  \includegraphics[width=.25\hsize]{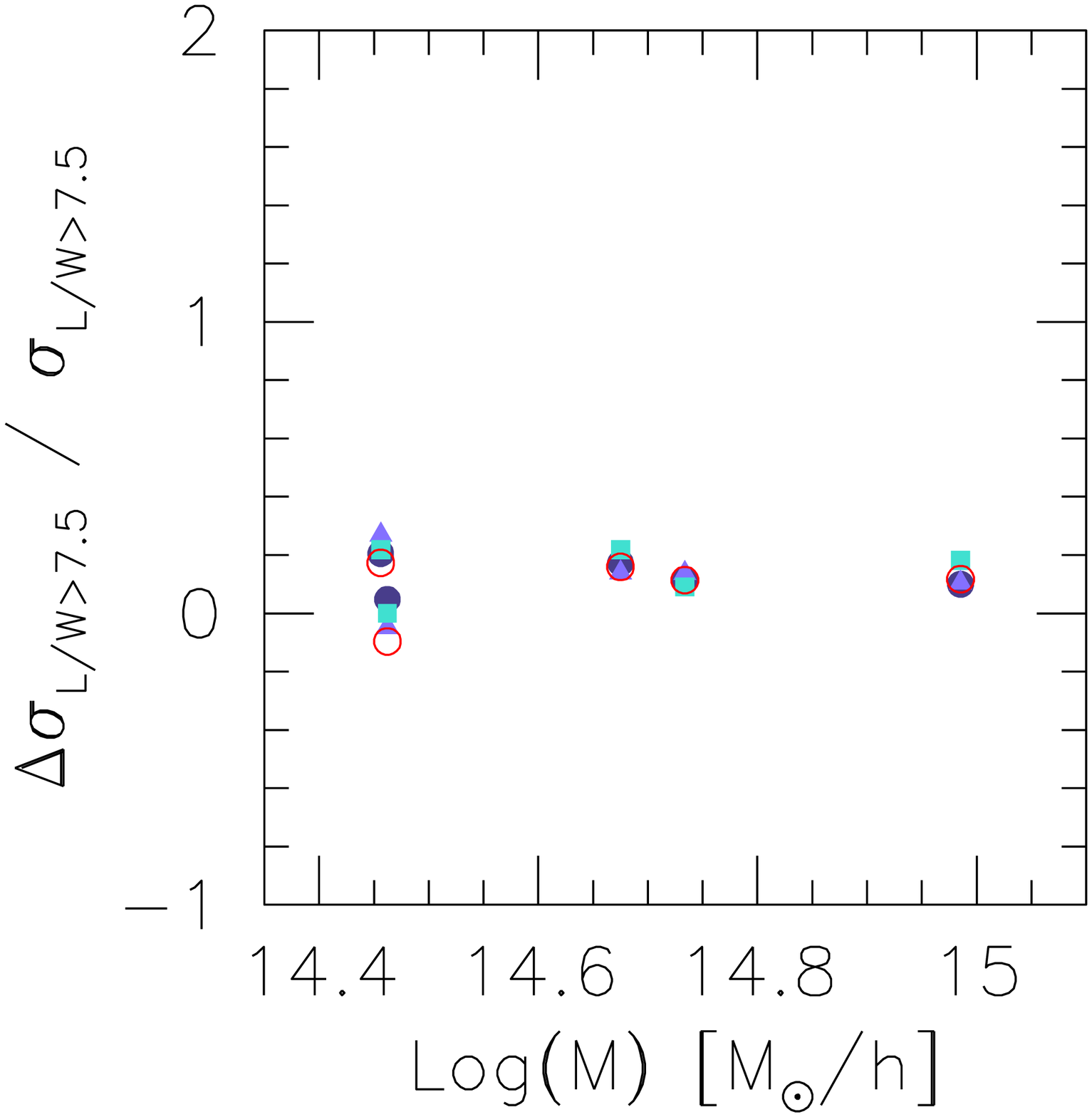}
  \includegraphics[width=.25\hsize]{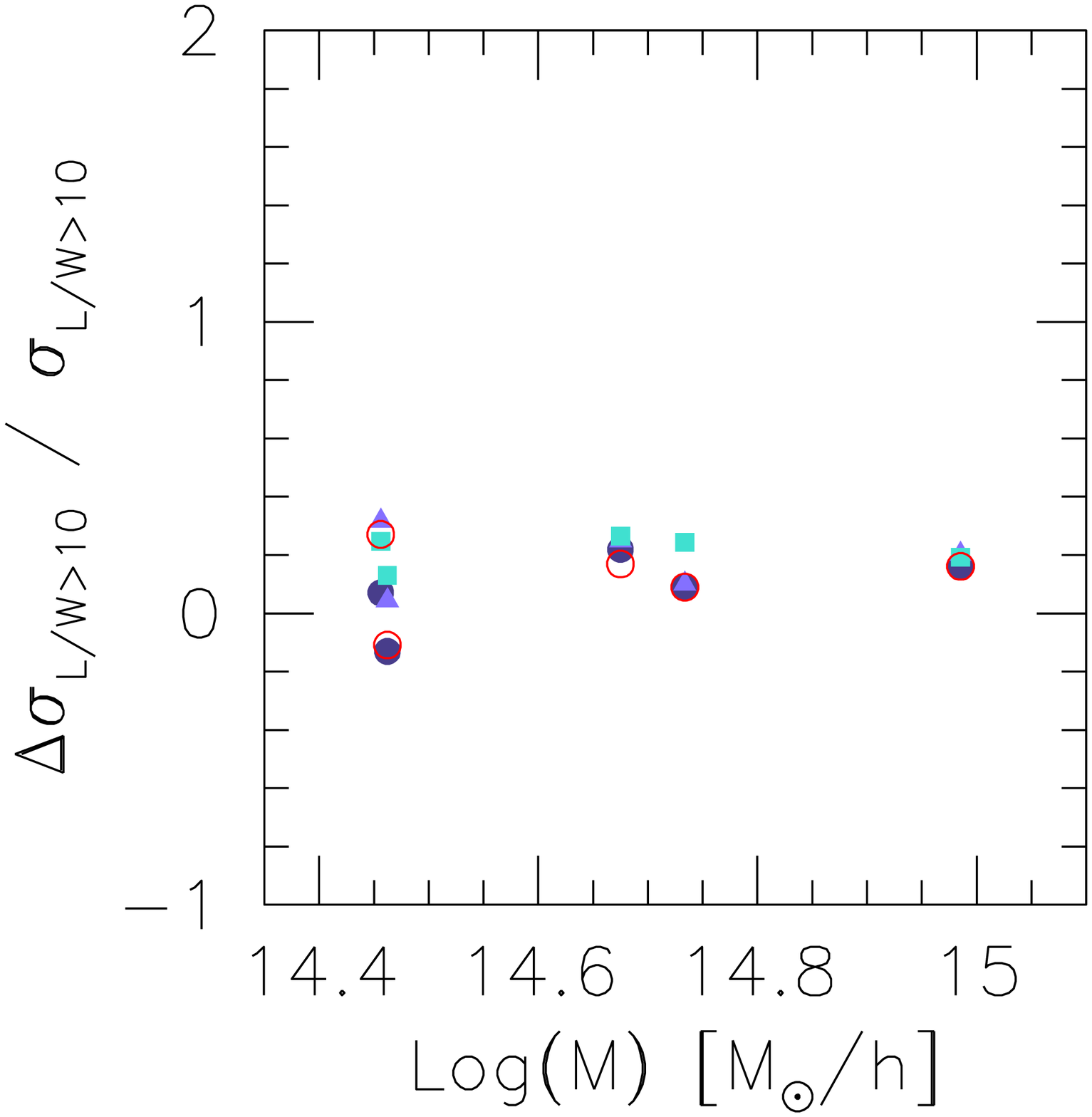}
  \includegraphics[width=.25\hsize]{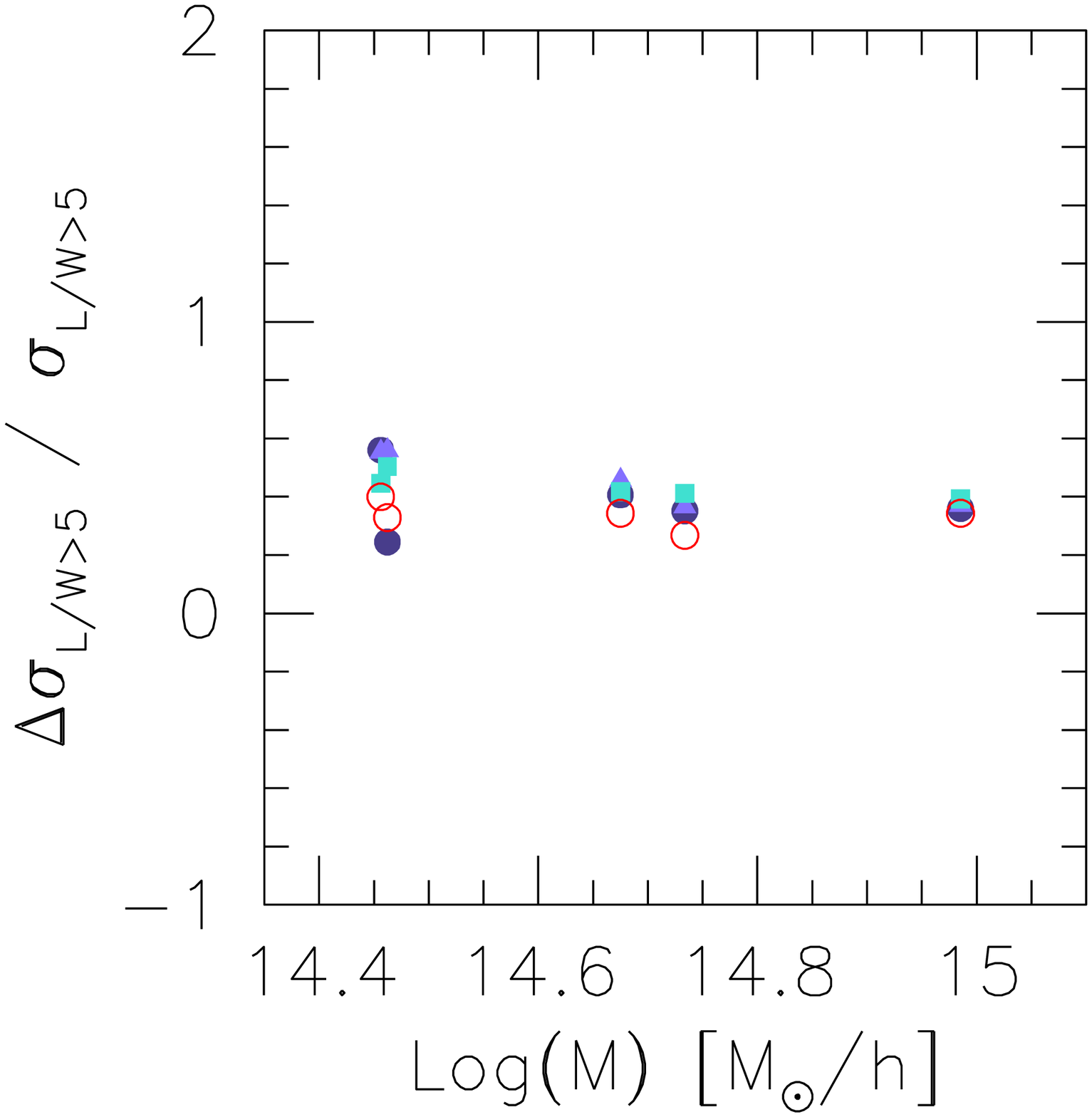}
  \includegraphics[width=.25\hsize]{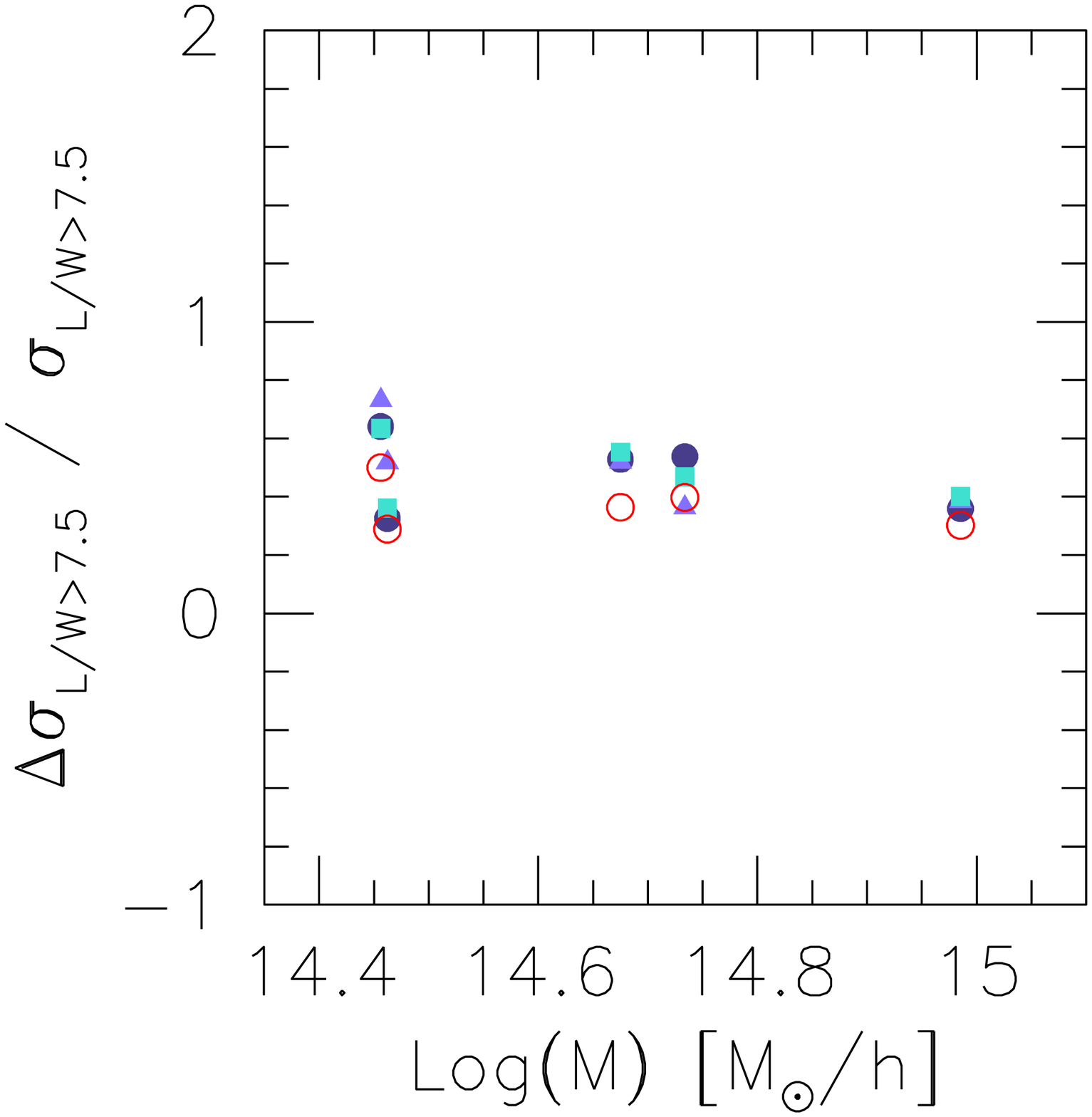}
  \includegraphics[width=.25\hsize]{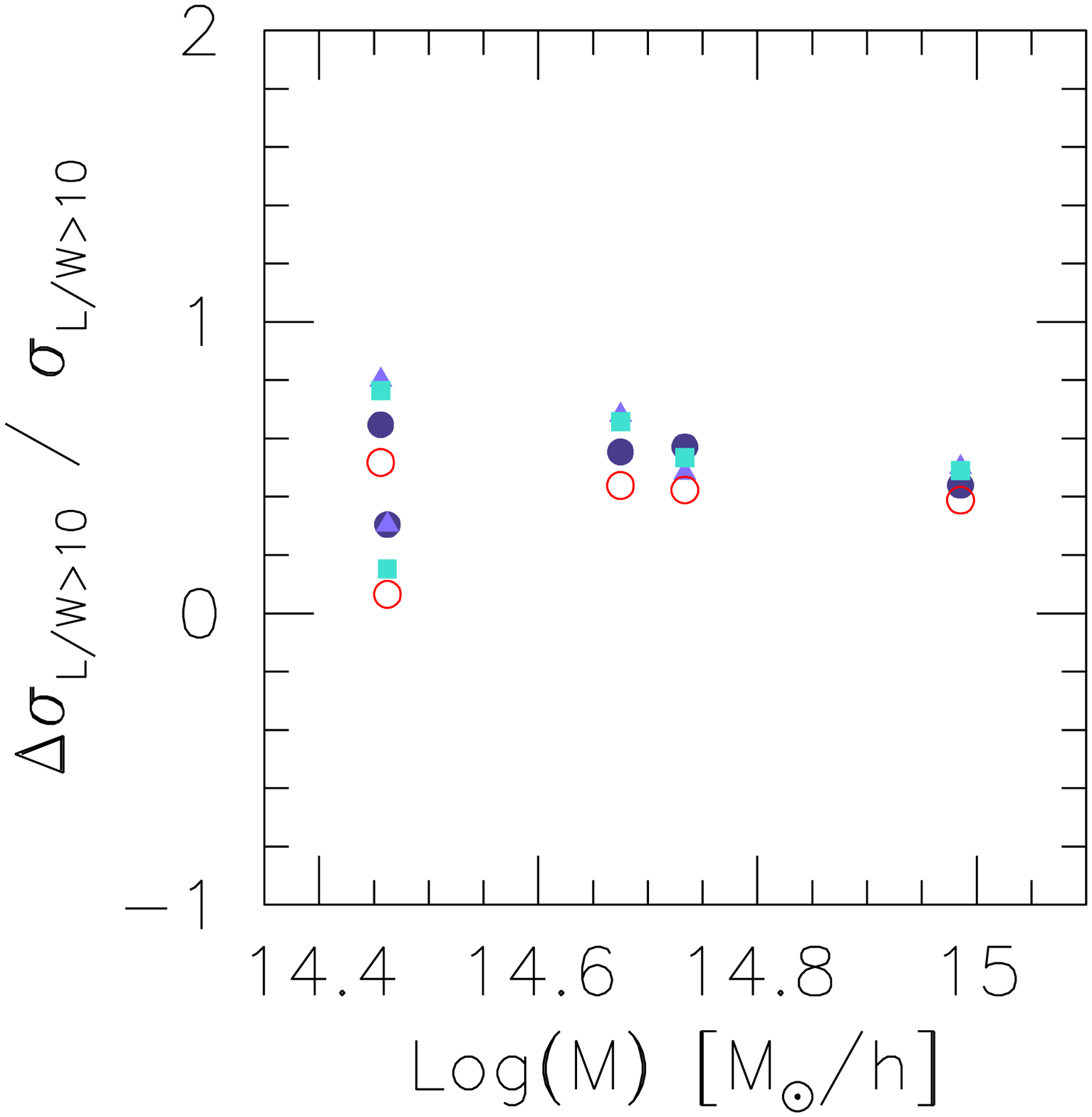}
\caption{Relative change in the cross sections for arcs with
  length-to-width ratios exceeding $5$ (left column), $7.5$ (central
  column) and $10$ (right column) as a function of cluster mass for
  the numerically simulated clusters in the OCDM model. Results are
  shown for three different masses of the cD galaxy:
  $5\times10^{12}\,h^{-1}\,M_\odot$ (top panels),
  $10^{13}\,h^{-1}\,M_\odot$ (middle panels) and
  $5\times10^{13}\,h^{-1}\,M_\odot$ (bottom panels). Circles, squares
  and triangles identify different cD galaxy models with the same
  symbols as in Fig.~(\ref{figure:crosssec_lambda}).}
\label{figure:crosssec_open}
\end{figure*}

Tangential arcs are selected from the catalogues of simulated images
for each numerical cluster model by requiring a minimal
length-to-width ratio. We then quantify the cluster's efficiency for
producing this type of arcs by measuring its lensing cross
section. This procedure is carried out for all numerical clusters
before and after the inclusion of the cD galaxy.

By definition, the lensing cross section is the area on the source
plane where a source must be located in order to be imaged as an arc
with the specified property. As explained in
Sect.~\ref{section:lenssim}, each source is taken to represent a
fraction of the source plane. We assign a statistical weight of unity
to the sources which are placed on the sub-grid with the highest
resolution. These cells have area $A$. The lensing cross section is
then measured by counting the statistical weights of the sources whose
images satisfy a specified property. If a source has multiple images
with the required characteristics, its statistical weight is
multiplied by the number of such images. Thus, the formula for
computing cross sections for arcs with a property $p$ is
\begin{equation}
  \sigma_\mathrm{p}=A\,\sum_i\,w_i\,n_i\;,
\end{equation}
where $n_i$ is the number of images of the
$i$-th source satisfying the required conditions, and $w_i$ is the
statistical weight of the source. 

We compute the lensing cross sections for arcs whose length-to-width
ratio $(L/W)$ exceeds a lower threshold $(L/W)_\mathrm{min}$. The
results obtained for each cluster with and without the cD galaxy are
compared by computing the relative change of the cross sections due to
the presence of the cD.
 
We show in Figs.~(\ref{figure:crosssec_lambda}) and
(\ref{figure:crosssec_open}) the relative change of the lensing cross
sections for long and thin arcs as a function of the cluster virial
mass for the simulations in the $\Lambda$CDM and in the OCDM models.
The panels in the left, central and right columns show the relative
change of the cross sections for arcs with length-to-width ratios
$(L/W)>5$, $(L/W)>7.5$ and $(L/W)>10$, respectively. In each panel, we
plot the results for all four models used to describe the cD
galaxy. The top, middle and bottom panels refer to simulations in
which the virial mass of the cD galaxy is $(5\times10^{12}, 10^{13},
5\times10^{13})\,h^{-1}\,M_\odot$, respectively.
 
As expected, the largest variations of the cross sections are
typically found if the cD galaxy is modelled as a pseudo-elliptical
NFW model whose orientation is aligned with that the host cluster. On
the other hand, cD galaxies with SIS profiles change the ability of
the numerical clusters for producing long and thin arcs only by a very
small amount.

In the $\Lambda$CDM model, a cD galaxy with mass between
$M_\mathrm{cD}=5\times10^{12}\,h^{-1}\,M_\odot$ and
$M_\mathrm{cD}=10^{13}\,h^{-1}\,M_\odot$ produces maximal variations
on the order of $40\%-50\%$. More massive cDs have a stronger impact
on the strong-lensing efficiency of the clusters: a galaxy with mass
$M_\mathrm{cD}=5\times10^{13}\,h^{-1}\,M_\odot$ changes the lensing
cross section by a maximum amount between $60\%$ and $200\%$,
depending on the total cluster mass. The impact of the cD is generally
larger in the less massive clusters.

A similar trend is found in the OCDM model, but the variations of the
cross sections are smaller in this case. For example, in the
simulations with the most massive cDs, the cross sections change by
approximately $40\%-80\%$ only. This behaviour was expected because
the clusters in the OCDM model are generally more compact compared
than those in the $\Lambda$CDM model. Including the cD, the mass in
the very central part of the clusters changes less compared to the
clusters in the $\Lambda$CDM model. Cross sections for arcs with
different minimal length-to-width ratios show similar variations.

\subsection{Radial arcs}  

\begin{figure*}\ifjournal\else[ht]\fi
\centering
  \includegraphics[width=.4\hsize]{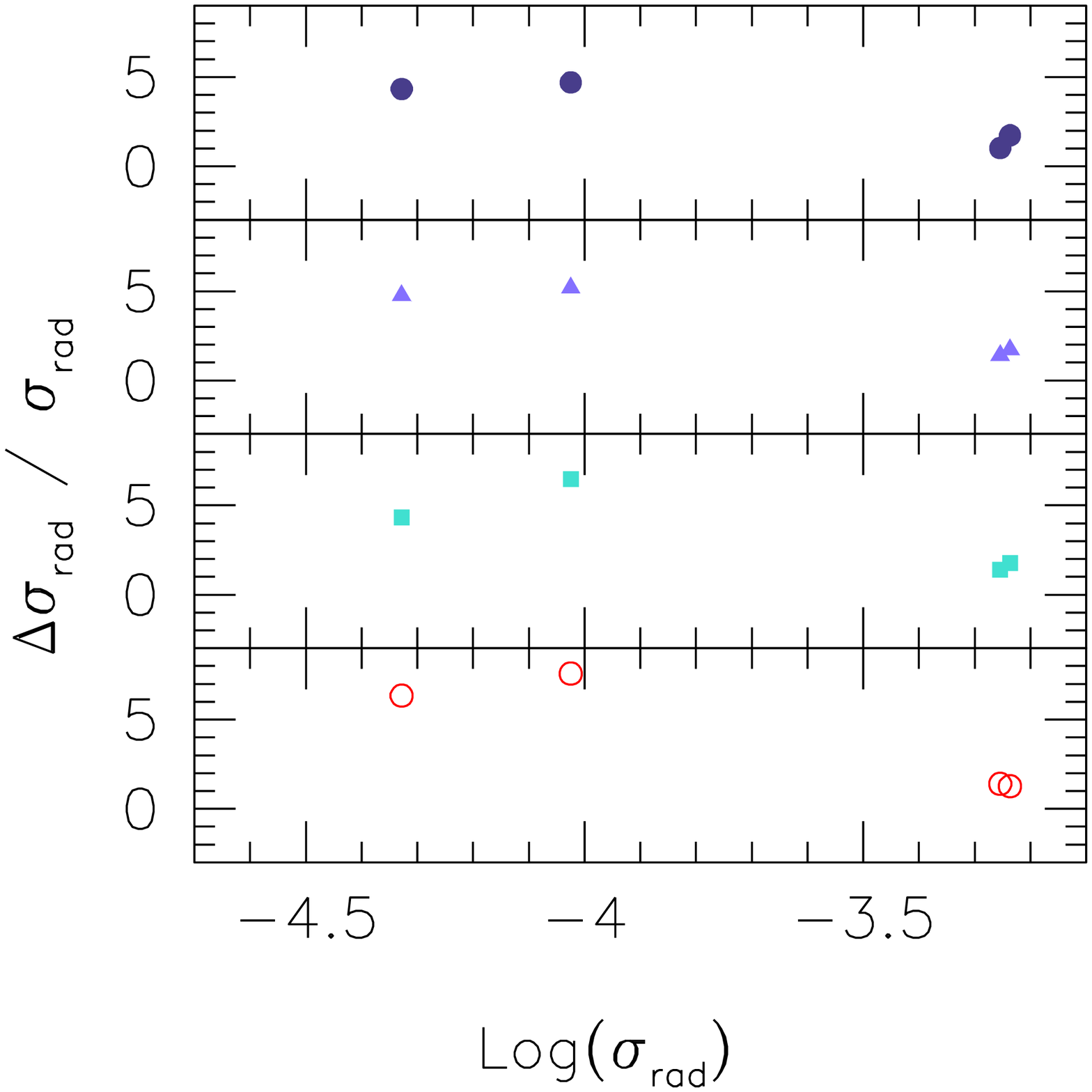}
  \includegraphics[width=.4\hsize]{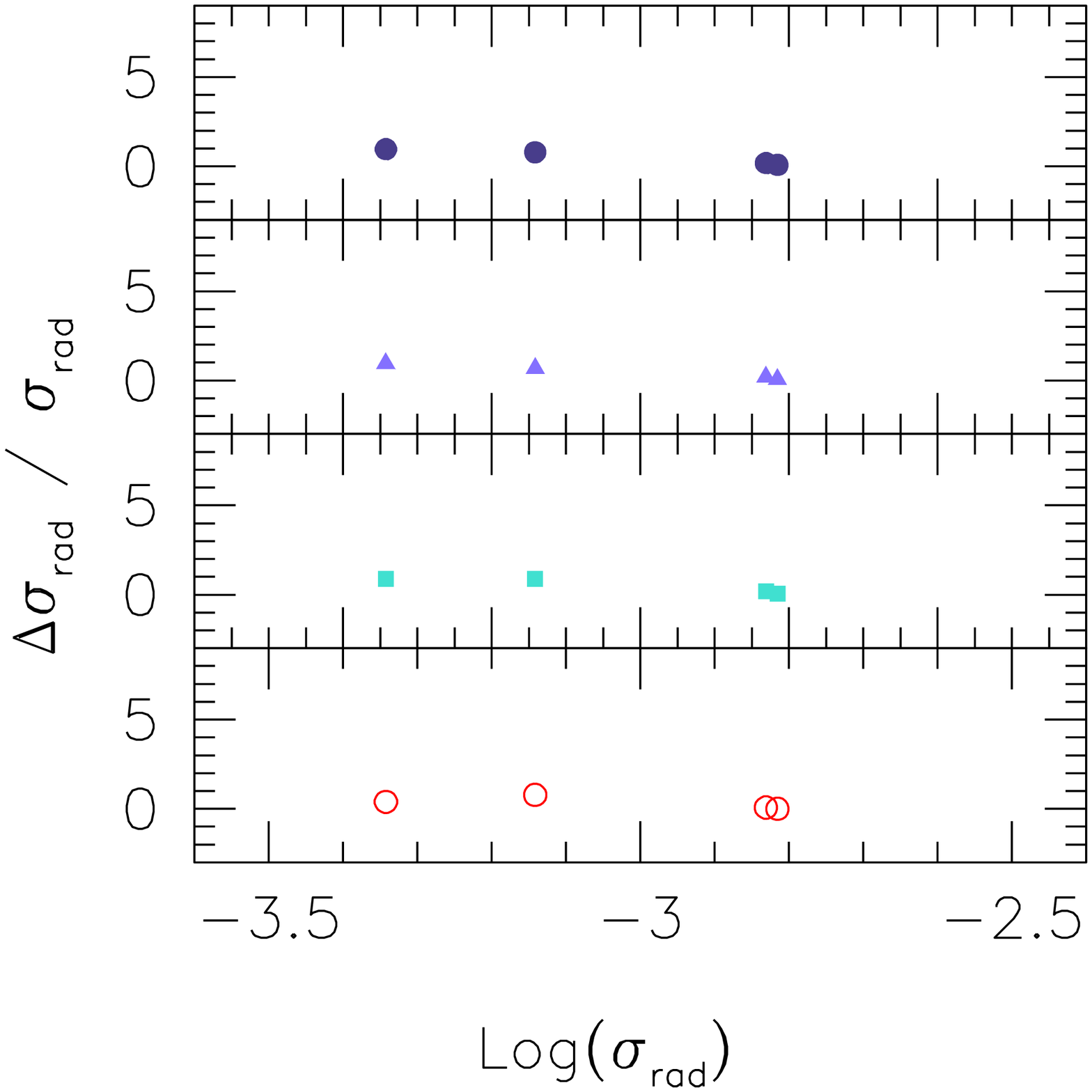} 
\caption{Relative change of the cross section for radial arcs versus
  original cross sections (i.e.~before the inclusion of a cD
  galaxy). The left and right panels show the results obtained in the
  $\Lambda$CDM and the OCDM models, respectively. We use circles,
  squares and triangles to identify different cD galaxy models, using
  the same symbols as in Figs.~(\ref{figure:crosssec_lambda}) and
  (\ref{figure:crosssec_open}).}
\label{figure:radialarcs}
\end{figure*}

The appearance and location of radial arcs within the lensing clusters
is determined by several factors, among them the slope of the
projected density profile and the central density of the lens. The
steeper the density profile is, the closer to the centre the radial
arcs move. Moreover, the higher the central surface density is, the
larger is the extent of the radial critical line.

Radial arcs are quite rare events. Many clusters in our sample exhibit
very short radial critical curves and are thus not very efficient in
producing radially distorted images. In particular, the numerical
clusters in our sample have significantly smaller cross sections for
radial arcs in the $\Lambda$CDM than in the OCDM model. The reason is
that the clusters in the $\Lambda$CDM model are less concentrated than
in the OCDM model. Given that we increase the mass in the inner part
of the numerical clusters by quite a large amount by adding a massive
cD galaxy, we expect that the capability of the cluster models for
producing radial arcs should be increased when a cD galaxy is
included. For investigating this question, we again compare the
lensing cross sections for radial arcs of the numerical models before
and after including a cD galaxy.

Radial arcs are identified from the complete sample of distorted
images using the technique described by Meneghetti et al.~(2001). It
consists in selecting those arcs for which the measured radial
magnification at their position exceeds a lower threshold. The
relative change of the lensing cross section for this type of arcs as
a function of the original lensing cross section of clusters not
containing a cD galaxy is shown in Fig.~(\ref{figure:radialarcs}). The
cD galaxy mass is $5\times10^{13}\,h^{-1}\,M_\odot$. In the left and
right panels, we plot the results for the cluster models in the
$\Lambda$CDM and in the OCDM models, respectively. Again we use
circles, squares and triangles for identifying different cD galaxy
models, using the same symbols as in
Figs.~(\ref{figure:crosssec_lambda}) and (\ref{figure:crosssec_open}).

As the figure shows, the cD galaxy increases the capability of the
cluster models for producing radial arcs, but this enhancement depends
on the original size of their lensing cross sections. Clusters which
were less efficient in producing radial arcs increase their lensing
cross sections by roughly $500\%$, while this increment is smaller for
lenses which already produced a higher number of radial arcs. In
particular, the effect of the cD galaxy on the lensing cross sections
of clusters is significantly larger in the $\Lambda$CDM model compared
to the OCDM model. One of the clusters in our sample (the least
massive one) did not produce any radial arcs before a cD galaxy was
included, thus the results for only four clusters are shown in
Fig.~(\ref{figure:radialarcs}).

\section{Summary and Conclusions}
\label{section:conclusions}

We investigated the effect of massive cD galaxies on the
strong-lensing properties of galaxy clusters. The main motivation
behind our study is to identify possible reasons for the pronounced
failure of numerical cluster-lensing models in the strongly favoured
$\Lambda$CDM cosmology in reproducing the observed abundance of giant
luminous arcs.

Although inadequate for quantitatively reliable results, our analytic
model demonstrated three interesting aspects of the problem. First,
the effect of cD galaxies is much stronger in clusters with NFW rather
than singular isothermal density profile because of the flatter central
density profile of the former. Second, cD galaxies are expected to be
much less efficient in asymmetric than in axially symmetric clusters
because the stronger gravitational tidal field in asymmetric clusters
makes the relative contribution of the additional central mass
component less important. Third, the impact of cD galaxies on the
strong-lensing properties of clusters is expected to be almost
independent of lens redshift.

We then carried out ray-tracing simulations using massive numerically
simulated galaxy clusters in the $\Lambda$CDM and OCDM cosmologies to
which we added cD galaxies in four different ways; either modelled as
singular isothermal spheres or modelled with the NFW density profile,
either axially symmetric or elliptically distorted; and if distorted,
either oriented randomly or aligned with the orientation of the
cluster mass distribution. All cD galaxies are placed at the minima of
the deflection-angle maps of the cluster models. They have masses of
$5\times10^{12}$, $10^{13}$, or $5\times10^{13}\,h^{-1}\,M_\odot$,
which are added to the cluster mass.

Note that these choices yield conservative results because they tend
to exaggerate the possible effects. The cD-galaxy masses are
\emph{very} high, in particular compared to the least massive cluster
of our sample, and they are \emph{added} to the clusters without
correspondingly decreasing the cluster mass.

Although we find relative enhancements of large-arc cross sections by
up to $\sim200\%$ for our least massive cluster if a cD galaxy with
$5\times10^{13}\,h^{-1}\,M_\odot$ and an elliptically distorted NFW
density profile is added with its orientation aligned with the mass
distribution of the cluster, such a situation can hardly be considered
realistic because that cD galaxy has $\sim20\%$ of the cluster
mass. In more realistic cases of cD galaxies with
$\lesssim10^{13}\,h^{-1}\,M_\odot$, tangential-arc cross sections are
increased by not more than $\sim50\%$. It should be noted, however,
that we have to compare the relative increase between clusters in
$\Lambda$CDM and OCDM models, which is typically not more than
$\sim30\%$.

Cross sections for the formation of radial arcs are much more affected
because they are highly sensitive to the exact central density profile
of the cluster lenses. The presence of a cD galaxy can multiply the
radial-arc cross sections by factors of a few, typically up to five in
the $\Lambda$CDM model. This reinforces the finding by other authors
that radial-arc statistics potentially is a highly significant tracer
for central density profiles of clusters, but as such less suitable
for constraints on cosmological parameters. Currently, also, the data
base for radial arc statistics is rather too poor for any meaningful
conclusions. In fact only few radial arcs have been discovered so far
and all of them have been found in clusters which are dominated by a
cD galaxy (e.g.~AC~114, MS~2137; Natarajan et al.~1998, Hammer et
al.~1997) or by one or two giant ellipticals (e.g.~A~370; B\'ezecourt
et al.~1999).

We thus conclude from our conservative estimates of the impact of cD
galaxies on strong-lensing cross sections by galaxy clusters that they
may increase the arc-formation probability by perhaps up to $\sim50\%$
in realistic situations, but certainly by far not enough for
explaining the discrepancy between simulations in $\Lambda$CDM models
and the observed abundance of arcs.

This result supports the view that we are still a long way from
understanding what exactly makes real clusters efficient strong
lenses. Numerous observational effects need to be taken into account
in addition to careful numerical models before detailed comparisons
between observed and simulated arcs can be carried out. The recent
suggestion by Wambsganss et al.~(2003) that arc statistics in
$\Lambda$CDM models can be reconciled with the observed frequency of
arcs by adopting a broader source redshift distribution is certainly
no solution either because the arc statistics problem as posed by
Bartelmann et al.~(1998) concerns the ability of a well-defined sample
of massive clusters to form arcs from a photometrically limited sample
of sources, rather than the strong-lensing optical depth of a section
of the Universe for arbitrarily faint sources. It is very likely that
careful comparisons between numerically simulated and observed
strong-lensing clusters will tell us much about the detailed mass
distribution and the history of mass assembly in individual clusters.

\section*{Acknowledgements}

This work has been partially supported by Italian MIUR (Grant 2001,
prot. 2001028932, ``Clusters and groups of galaxies: the interplay of
dark and baryonic matter''), CNR and ASI.  MM
thanks the EARA for financial support and the Max-Planck-Institut
f\"ur Astrophysik for the hospitality during the visits when part of
this work was done.  We are grateful to Bepi Tormen for clarifying
discussions.


\begin{thebibliography}{}
\bibitem[\protect\citename{Austin \& Peach} 1974]{austin74} Austin T.~B.,
Peach J.~V., 1974, MNRAS,  168, 591
\bibitem[\protect\citename{Bartelmann} 1996]{bartelmann96} Bartelmann
M., 1996, A\&A,  313, 697 
\bibitem[\protect\citename{Bartelmann et al.} 1998]{bartelmann98}
Bartelmann M., Huss A., Colberg J.~M., Jenkins A., Pearce F.~R., 1998,
A\&A, 330, 1
\bibitem[\protect\citename{Bartelmann et al.} 2003]{bartelmann03}
Bartelmann M., Meneghetti M., Perrotta F., Baccigalupi C., Moscardini L., 2003,
preprint, astro-ph/0210066 
\bibitem[\protect\citename{Bartelmann \& Weiss }1994]{bartelmann94}
Bartelmann M., Weiss A., 1994, A\&A, 298, 1
\bibitem[\protect\citename{B{\' e}zecourt} 1999]{bezecourt99} B{\'
e}zecourt J., Kneib J.~P., Soucail G., Ebbels T.~M.~D., 1999, A\&A,  347,
21 
\bibitem[\protect\citename{Binggeli} 1982]{binggeli82} Binggeli
B., 1982, A\&A,  107, 338
\bibitem[\protect\citename{Bond \& Efstathiou }1984]{bond84} Bond
  J.R., Efstathiou G., 1984, ApJ, 285, L45
\bibitem[\protect\citename{Bullock et al.} 2001]{bullock01} Bullock
J.~S., Kolatt T.~S., Sigad Y., Somerville R.~S., Kravtsov A.~V., Klypin
A.~A., Primack J.~R., Dekel A., 2001, MNRAS,  321, 559 
\bibitem[\protect\citename{Carter} 1980]{carter80} Carter D.,
1980, MNRAS,  190, 307
\bibitem[\protect\citename{Cooray, Quashnock \& Miller} 1999]{cooray99} Cooray A.~R.,
Quashnock J.~M., Miller M.~C., 1999, ApJ,  511, 562
\bibitem[\protect\citename{de Bernardis et al.} 2002]{debernardis02} de
Bernardis P., Ade P.~A.~R., Bock J.~J., et al., 2002, ApJ,  564, 559
\bibitem[\protect\citename{Eke et al.} 2001]{eke01} Eke V.~R.,
Navarro J.~F., Steinmetz M., 2001, ApJ,  554, 114
\bibitem[\protect\citename{Garijo et al.} 1997]{garijo97} Garijo A.,
Athanassoula E., Garcia-Gomez C., 1997, A\&A,  327, 930
\bibitem[\protect\citename{Gioia \& Luppino} 1994]{gioia94} Gioia I.~M.,
Luppino G.~A., 1994, ApJS,  94, 583
\bibitem[\protect\citename{Gladders et al.} 2003]{gladders03}
Gladders M.D., Hoekstra H., Yee H.K.C., Hall P.B., Barrientos L.F., 2003,
preprint, astro-ph/0303341 
\bibitem[\protect\citename{Hammer} 1997]{hammer97} Hammer F.,
Gioia I.~M., Shaya E.~J., Teyssandier P., Le Fevre O., Luppino G.~A., 1997,
ApJ,  491, 477 
\bibitem[\protect\citename{Hockney \& Eastwood }1988]{hockney88}
Hockney R.W., Eastwood J.W., 1988, Computer simulations using
particles. Hilger, Bristol
\bibitem[\protect\citename{Kauffmann et al. }1999]{kauffmann99}
  Kauffmann G.A.M., Colberg J.M., Diaferio A., White S.D.M., 1999,
  MNRAS, 303, 188
\bibitem[\protect\citename{Kaufmann \& Straumann }2000]{kaufmann00}
  Kaufmann R., Straumann N., 2000, Ann. Phys., 11, 507
\bibitem[\protect\citename{Lambas et al.} 1988]{lambas88} Lambas D.~G.,
Groth E.~J., Peebles P.~J.~E., 1988, AJ,  95, 975
\bibitem[\protect\citename{Le Fevre et al.} 1994]{lefevre94} Le Fevre
O., Hammer F., Angonin M.~C., Gioia I.~M., Luppino G.~A., 1994, ApJ,  422,
L5
\bibitem[\protect\citename{Luppino et al.} 1999]{luppino99} Luppino
G.~A., Gioia I.~M., Hammer F., Le F{\` e}vre O., Annis J.~A., 1999, A\&AS,
136, 117
\bibitem[\protect\citename{Meneghetti et al. }2003]{meneghetti02}
  Meneghetti M., Bartelmann M., Moscardini L., 2003, MNRAS, 340, 105
\bibitem[\protect\citename{Meneghetti et al. }2000]{meneghetti00}
Meneghetti M., Bolzonella M., Bartelmann M., Moscardini L., Tormen G.,
2000, MNRAS, 314, 338 
\bibitem[\protect\citename{Meneghetti et al. }2001]{meneghetti01}
  Meneghetti M., Yoshida N., Bartelmann M., Moscardini L., Springel
  V., Tormen G., White S.D.M., 2001, MNRAS, 325, 435
\bibitem[\protect\citename{Molikawa \& Hattori }2001]{molikawa01}
Molikawa K., Hattori M., 2001, ApJ, 559, 544
\bibitem[\protect\citename{Molikawa et al. }1999]{molikawa99}
Molikawa K., Hattori M., Kneib J.-P., Yamashita K., 1999, A\&A, 351,
413
\bibitem[\protect\citename{Natarajan} 1998]{natarjan98} Natarajan
P., Kneib J., Smail I., Ellis R.~S., 1998, ApJ,  499, 600
\bibitem[\protect\citename{Navarro et al. } 1996]{navarro96} Navarro
J.~F., Frenk C.~S., White S.~D.~M., 1996, ApJ,  462, 563
\bibitem[\protect\citename{Navarro et al. }1997]{navarro97} Navarro
J.F., Frenk C.S., White S.D.M., 1997, ApJ, 490, 493
\bibitem[\protect\citename{Oguri et al. }2001]{oguri01} Oguri M.,
  Taraya A., Suto Y., 2001, ApJ, 559, 572
\bibitem[\protect\citename{Oguri et al. }2002]{oguri02} Oguri M.,
  Taraya A., Suto Y., Turner E.L., 2002, ApJ, 568, 488
\bibitem[\protect\citename{Perlmutter} 1998]{perlmutter98}
Perlmutter S., 1998, AAS,  30, 1388
\bibitem[\protect\citename{Perlmutter et al. } 1997]{perlmutter97}
Perlmutter S., Gabi S., Goldhaber G., et al., 1997, ApJ,  483, 565 
\bibitem[\protect\citename{Perrotta et al. }2002]{perrotta01} Perrotta
  F., Baccigalupi C., Bartelmann M., de Zotti G., Granato G.L.,
 2002, MNRAS, 329, 445
\bibitem[\protect\citename{Porter et al. } 1991]{porter91} Porter A.~C.,
Schneider D.~P., Hoessel J.~G., 1991, AJ,  101, 1561
\bibitem[\protect\citename{Rhee \& Katgert} 1987]{rhee87} Rhee
G.~F.~R.~N., Katgert P., 1987, A\&A,  183, 217
\bibitem[\protect\citename{Rood \& Sastry} 1972]{rood72} Rood H.~J.,
Sastry G.~N., 1972, AJ,  77, 451
\bibitem[\protect\citename{Sastry} 1968]{sastry68} Sastry G.~N.,
1968, PASP,  80, 252
\bibitem[\protect\citename{Struble \& Peebles } 1985]{struble85} Struble
M.~F., Peebles P.~J.~E., 1985, AJ,  90, 582
\bibitem[\protect\citename{Torri et al. }2003]{torri03}
Torri E., Meneghetti M., Bartelmann M., Moscardini L., Rasia E.,
Tormen G., 2003, preprint
\bibitem[\protect\citename{Viana \& Liddle }1996]{viana96}
Viana P.T.P., Liddle A.R., 1996, MNRAS, 281, 323
\bibitem[\protect\citename{Wu \& Hammer } 1993]{wu93} Wu X., Hammer F.,
1993, MNRAS,  262, 187 
\bibitem[\protect\citename{Wambsganss et al. } 2003]{wbo03}
Wambsganss, J., Bode, P., Ostriker, J.P., 2003, ApJ submitted;
preprint astro-ph/0306088
\bibitem[\protect\citename{Wu \& Mao }1996]{wu96}
Wu X.-P., Mao S., 1996, ApJ, 463, 404
\bibitem[\protect\citename{Zaritsky \& Gonzalez} 2003]{zaritsky03}
Zaritsky D., Gonzalez A.~H., 2003, ApJ,  584, 691 
\end{thebibliography}
\end{document}